%% file: main.tex
\documentclass[conference]{IEEEtran}

\newif\ifAnon\Anonfalse

\usepackage[hyphens]{url}
\PassOptionsToPackage{bookmarks,hidelinks}{hyperref}
\usepackage{hyperref}
\hypersetup{breaklinks=true}
\usepackage{tugraz_defaults}

\usepackage{csquotes}
\usepackage{longtable}

\usepackage{graphicx}
\usepackage{caption} 
\usepackage{subcaption}
\usepackage{pgfplotstable}
\usepackage{hyperref}

\renewcommand{\paragraph}[1]{\vspace{0.0cm}\noindent\textbf{#1}\ }

\usepgfplotslibrary{dateplot}
\usepgfplotslibrary{fillbetween}

\makeatletter
    \newenvironment{myalign*}{%
      \setlength{\abovedisplayskip}{-0.2\baselineskip}%
      \setlength{\abovedisplayshortskip}{\abovedisplayskip}%
      \start@align\@ne\st@rredtrue\m@ne
    }%
    {\endalign}
\makeatother

\graphicspath{{figures/}}

\usepackage[framemethod=tikz]{mdframed}
\mdfdefinestyle{mystyle}{%
  rightline=true,
  innerleftmargin=5,
  innerrightmargin=5,
  innertopmargin=0,
  outerlinewidth=0.5pt,
  topline=true,
  rightline=true,
  bottomline=true,
  roundcorner=1pt,
  skipabove=0.6\topskip,
  skipbelow=0.2\topskip
}

\mdfdefinestyle{tightstyle}{%
  rightline=true,
  innerleftmargin=2,
  innerrightmargin=2,
  innerbottommargin=0,
  innertopmargin=0,
  outerlinewidth=0.3pt,
  topline=true,
  rightline=true,
  bottomline=true,
  roundcorner=1pt,
  skipabove=0.6\topskip,
  skipbelow=0.2\topskip
}

\newcommand{\paragrabf}[1]{\noindent \textbf{#1}\ }

\usepgfplotslibrary{dateplot}
\usepgfplotslibrary{fillbetween}

\newcommand{\TechniqueName}{\emph{LBTA}\xspace}
\newcommand{\TechniqueNames}{\emph{LBTAs}\xspace}
\newcommand{\FullAttackName}{\emph{Layered Binary Templating Attack}}

\newif\ifShortVersion\ShortVersiontrue

\usepackage{comment}

\begin{document}

\title{Layered Binary Templating: Efficient Detection of Compiler- and Linker-introduced Leakage}

\date{}

\ifAnon
  \author{}
\else
  \author{
  {\rm Martin Schwarzl}\\
  {Graz University of Technology}
  \and
  {\rm Erik Kraft}\\
  {Graz University of Technology}
  \and
  {\rm Daniel Gruss}\\
  {Graz University of Technology}
  }
\fi


\maketitle

\begin{abstract}
  Cache template attacks demonstrated automated leakage of user input in shared libraries.
  However, for large binaries, the runtime is prohibitively high.
  Other automated approaches focused on cryptographic implementations and media software but are not directly applicable to user input.
  Hence, discovering and eliminating all user input side-channel leakage on a cache-line granularity within huge code bases are impractical.

  In this paper, we present a new generic cache template attack technique, \TechniqueName, layered binary templating attacks.
  \TechniqueName uses multiple coarser-grained side channel layers as an extension to cache-line granularity templating to speed up the runtime of cache templating attacks.
  We describe \TechniqueName with a variable number of layers with concrete side channels of different granularity, ranging from \SI{64}{\byte} to \SI{2}{\mega\byte} in practice and in theory beyond.
  In particular the software-level page cache side channel in combination with the hardware-level L3 cache side channel, already reduces the templating runtime by three orders of magnitude.
  We apply \TechniqueNames to different software projects and thereby discover data deduplication and dead-stripping during compilation and linking as novel security issues.
  We show that these mechanisms introduce large spatial distances in binaries for data accessed during a keystroke, enabling reliable leakage of keystrokes.
  Using \TechniqueName on Chromium-based applications, we can build a full unprivileged cache-based keylogger\footnote{In the demo provided, the user first announces via Signal messenger to send money to a friend, then switches to Google Chrome to visit a banking website and enters the credentials there. All keystrokes are correctly leaked. \url{https://streamable.com/dgnuwk}.}.
  Our findings show that all user input to Chromium-based apps is affected and we demonstrate this on a selection of popular apps including Signal, Threema, Discord, and password manager apps like passky.
  As this is not a flaw of individual apps but the framework, we conclude that all apps that use the framework will also be affected, \ie hundreds of apps.

\end{abstract}

\section{Introduction}
Side-channel attacks are a powerful technique to leak information from side effects of computations~\cite{Kocher1996}.
Especially caches, buffering memory recently accessed, are a popular attack target, with generic attack techniques like \FlushReload enabling attacks with a high spatial and temporal resolution and high accuracy.
The first cache attacks were focused on cryptographic algorithms~\cite{Page2002,Tsunoo2003,Bernstein2005,Percival2005,Osvik2006,Lawson2009,Gullasch2011,Inci2015,Guelmezoglu2015,Gulmezoglu2016cross,Inci2016}.
In the last decade, the focus has been extended to non-cryptographic applications that still operate on secret data, \eg breaking address-space layout randomization~\cite{Hund2013,Gras2017aslr,Barresi2015,Dixon2017,Koschel2020,Kim2021Breaking}, attacking secure enclaves~\cite{Gotzfried2017,Brasser2017sgx,Moghimi2017,Schwarz2017MGX,Dall2018cachequote}, spying on websites and user input~\cite{Lipp2016,Schwarz2018KeyDrown,Wang2019Unveiling}, and covert channels~\cite{Xu2011,Yarom2014Flush,Lipp2016,Maurice2017Hello,Saileshwar2021Streamline}.
In particular, user input, especially keystrokes, has become a popular attack target for inter-keystroke timing attacks~\cite{Song2001,Ristenpart2009,Lipp2016,Naghibijouybari2018,Schwarz2018KeyDrown,Wang2019Unveiling,Paccagnella2021lotr}.
Gruss~\etal\cite{Gruss2015Template} demonstrated that some libraries might leak more information than just inter-keystroke timings, being able to distinguish groups of keys, \eg number and alphabetic keys.

Compilers can also introduce side-channel leakage, which is invisible on the source level, by applying various optimizations with respect to the program's runtime, memory footprint, and binary size.
Moreover, link-time optimizations~\cite{Moser2006Optimizing} can further optimize binaries in the linking stage.
Page~\cite{Page2006A} showed that dynamic compilation could lead to side-channel leakage in a cryptographic library with constant-time implementation.
Simon~\etal\cite{Simon2018What} demonstrated that different optimization levels in common C compilers can break constant-time implementation of cryptographic primitives by introducing timing side channels.
Brennan~\etal\cite{Brennan2020JIT} discovered that timing side channels can be introduced by exploiting JIT compilation.
Hence, side channels introduced in the development process due to the complex interaction between software components and compilers that are not obvious on the software layer may remain undetected and pose a risk for all computer users.

As a consequence, numerous works explored the automatic identification of cache side-channel leakage.
Most of these works focus on cryptographic implementations and aim for the goal of making the code constant-time~\cite{Chari2002template,Rechberger2004Practical,Medwed2008Template,Brumley2009}.
More recently, Yuan~\etal\cite{Yuan2021Automated} showed that manifold learning can be used for automated side-channel analysis of media software.
For common applications processing sensitive input, \eg browsers, the situation is less clear, as it is not feasible to linearize and unify the entire instruction stream for different user inputs that trigger vastly different program behavior.
Cryptographic implementations must be resistant to repeatable attack attempts, while other form of input, such as user input, is unpredictable and non-repeatable~\cite{Schwarz2018KeyDrown}.
Such attacks often suffer from a noisy channel, \eg affected by co-location with other data on the same cache line or prefetching effects due to spatial proximity.
Therefore, the state-of-the-art is not to analyze the leakage based on models and theoretical leakage assumptions but to actively use side channels to template (\ie \emph{to profile}) the leakage that can be observed~\cite{Gruss2015Template,Wang2019Unveiling}.
In a cache-template attack the attacker maps binaries as shared memory into the address space of the templating process and profiles which memory locations show side-channel activity upon specific events.
In the attack phase, the attacker also maps binaries as shared memory, as this requires no privileges beyond read access.
Thus, the two attack phases operate on binary offsets and are entirely unaffected by mechanisms such as ASLR (address-space layout randomization).
To achieve a high level of confidence, the templating (\ie \emph{profiling}) must be performed on an attacker-controlled system that is identical or very similar to the target system and using the side channel that will also be used for exploitation subsequently.
Unfortunately, this comes at a high cost, namely the runtime of the templating.
For instance, templating the binary with cache template attacks~\cite{Gruss2015Template}, shared libraries, and memory-mapped files used by the Chrome browser (about \SI{210}{\mega\byte}), would take \SIx{113.17} days with the published cache template attack tool~\cite{Gruss2015Template} on our test system.
This leads to an unsatisfying situation, where cache leakage analysis is not part of secure development workflows.

In this paper, we present \TechniqueName, \FullAttackName.
\TechniqueName uses multiple different layers with different spatial granularities as an extension to cache-line granularity templating.
While a smaller spatial granularity is beneficial in the attack phase, it comes at a substantial slowdown in the templating phase.
\TechniqueName solves this problem by introducing a previously unexplored dimension into software-based templating attacks.
\TechniqueName combines the information of multiple side channels that provide information, \eg at different spatial granularity, to optimize and accelerate the search for secret-dependent activity.
Our templating starts with the channel with the most coarse spatial granularity and based on the activity uses more fine-grained spatial granularity to detect the exact location (cache-line granularity \SI{64}{\byte}).

Our evaluation of \TechniqueName on state-of-the-art systems shows that a variety of hardware and software channels with different granularity are available.
We focus in particular on a combination of a software channel, the page-cache side channel, with \SI{4}{\kilo\byte} granularity, and the cache side channel, with \SI{64}{\byte} granularity.
Page cache attacks are hardware-agnostic~\cite{Gruss2019page}, resulting in cross-platform applicability, \ie our templater supports both Windows and Linux with the \SI{4}{\kilo\byte} page-cache side channel.
We show that this two-layered approach already speeds up cache templating~\cite{Gruss2015Template} by three orders of magnitude (\ie \SIx{1848}x).

We evaluate \TechniqueName on different large software projects, including Chrome, Firefox, and LibreOffice Writer.
The most significant finding is a novel security issue introduced by compiler and linker optimizations such as \textbf{data deduplication during compilation and linking} and \textbf{dead-stripping}.
Linker optimizations introduce a spatial distance of multiple \SI{4}{\kilo\byte} pages between key-dependent data accessed during a keystroke.
Using \TechniqueName~\cite{Statcounter2022Browser}, we find distinct leakage for all alphanumeric keys, allowing us to build a full unprivileged cache-based keylogger using \FlushReload that leaks all keystrokes from Chromium-based applications involving password input fields, \eg Google Chrome on banking websites, popular messengers including Signal, Threema, Discord, and password manager apps like passky.
Based on our findings, we conclude that multiple hundreds of apps using the Chromium framework are affected~\cite{Electron2022Apps}.
Using \FlushReload we can mount powerful keyloggers with high accuracy for Chromium-based applications.
In addition, we demonstrate keystroke-correlated cache activity on Firefox and LibreOffice Writer.
While the cache activity does not reveal the actual keystoke, inter-keystroke timing attacks might be mounted on these targets~\cite{Song2001,Ristenpart2009,Zhang2009keystroke,Diao2016,Gruss2015Template}.
We confirm that on Linux, the \texttt{preadv2} syscall can be used instead of the meanwhile mitigated \texttt{mincore} syscall~\cite{Gruss2019page}, to perform page cache attacks~\cite{Corbet2019Fixing}.
However, accurate keylogging using with \texttt{preadv2} and the page cache attacks are infeasible, as stable results require a delay of \SIx{2} seconds between cache activity.

Since existing system hardening techniques, \eg ASLR (address-space layout randomization), have no effect on our attack, we provide a systematic discussion of the possible mitigation vectors.

\paragrabf{Contributions.} The main contributions of this work are:
\begin{compactenum}
  \item We introduce a new dimension, side-channel granularity, into cache template attacks and exploit it to speed up the templating runtime by three orders of magnitude.
  \item We show that the leakage discovered by \TechniqueName can be exploited both in hardware (\ie \FlushReload) and software cache attacks (\ie page cache attacks).
  \item With \TechniqueName, we discover optimizations in LLVM/clang that introduce side-channel leakage invisible on the source level, namely data deduplication and dead-stripping.
  \item Our end-to-end keylogger attack can be applied to Chromium-based software, including Google Chrome, secure messengers like Signal, and password managers.
\end{compactenum}

\paragrabf{Responsible Disclosure.}
We responsibly disclosed our findings to the Chromium project.
The underlying issue was rated as `medium' severity and a CVE will be assigned at the beginning of August 2022.
We are working with the Chromium team on resolving the vulnerability and informing all affected parties.
In alignment with them, our experiments and tools will be open-sourced after the vulnerability is patched.
The patch is already implemented and will be included in the Chromium release M104 in the beginning of August 2022. 
\footnote{We will provide a link via the chairs as soon as patches are ready.}

\paragrabf{Outline.}
In \cref{sec:background}, we provide the background.
In \cref{sec:attack}, we explain the building blocks of \TechniqueName.
In \cref{sec:compiler_sc}, we describe the effects during compilation and linking that facilitate or even introduce side-channel leakage, \ie data deduplication and dead-stripping.
In \cref{sec:evaluation}, we evaluate \TechniqueName on different attack targets and show a full end-to-end keylogger attack on the Chrome browser and Electron apps.
In \cref{sec:mitigation}, we provide a systematic analysis of mitigation vectors.
In \cref{sec:discussion}, we discuss its implications before we conclude in \cref{sec:conclusion}.

\section{Background}\label{sec:background}
In this section, we explain the necessary preliminaries on hard- and software cache attacks, automated discovery of side channel attacks, and side channels introduced by compiler optimizations.

\subsection{Shared Memory}
Operating systems apply various optimizations to reduce the systems general memory footprint.
One such optimization is shared memory, where the operating system actively tries to remove duplicate data mappings.
An example would be shared libraries, such as the \texttt{glibc}, which is used in many programs, and thus, can be shared between processes.
Moreover, with the \texttt{mmap} respectively \texttt{LoadLibrary} functions, a user program can request shared memory from the operating system by mapping the library as \texttt{read-only} memory.
Another optimization to reduce the memory footprint commonly used for virtual machines is memory deduplication on a pagewise level.
The operating system deduplicates pages with identical content and maps the deduplicated page in a copy-on-write semantic.

\subsection{Deduplication}\label{sec:dedup}
The concept of deduplication is very generic and can be applied in the context of various memory systems to save memory.
For storage systems, one example is cloud storage systems that deduplicate files to minimize the amount of storage required~\cite{Harnik2010,Keelveedhi2013dupless}.
For main memory, there are multiple mechanisms that lead to deduplicated memory.
One is copy-on-write, which avoids duplicating memory in the first place during process creation.
A similar effect can be achieved via the operating system's page cache~\cite{Gruss2019page}, \ie file-based page deduplication.
Modern operating systems also deduplicate the zero page, \ie many processes can map the same zero page as copy-on-write instead of having their own copies of zeroed memory.
However, the most prominent example is data-based page deduplication~\cite{Suzaki2011}.
With data-based page deduplication, the operating system or hypervisor scans the main memory page-wise and identifies identical pages, \eg using hashes or byte-wise data comparison.
The identical pages are then deduplicated and marked as copy-on-write.
In all cases of deduplication, when a user tries to modify the deduplicated memory, it is copied (duplicated) again.
All above types of deduplication are known to introduce side-channel leakage, \eg file deduplication~\cite{Harnik2010,bacs2022dupefs}, and also page deduplication~\cite{Suzaki2011} also from JavaScript~\cite{Gruss2015dedup,Costi2022On} and even remotely~\cite{Schwarzl2022Remote}.
Deduplication has also been a building block for other attacks such as KASLR breaks~\cite{Xiao2013security}, and Rowhammer attacks~\cite{Bosman2016,Razavi2016}.

While all above types of deduplication target memory, there is also deduplication in other contexts.
In this paper, we focus on a different type of deduplication that has little to do with the above or memory systems in general.
We instead focus on deduplication during compilation and linking.
The goal of deduplication here is similar, \ie reducing memory usage.
However, the security implications of deduplication during compilation and linking have not been studied so far.

\subsection{Cache Attacks}
Caches are fast and reliable hardware buffers used to speed up the memory access times.
There is a significant timing difference between cached and uncached data, which leads to a powerful cache side channel.
Kocher~\cite{Kocher1996} demonstrated that timing attacks on cryptographic primitives are possible \ie via caches or non constant-time arithmetic operations.
Cache attacks were used to attack cryptographic primitives~\cite{Kocher1996,Page2002,Tsunoo2003,Bernstein2005,Percival2005,Osvik2006,Lawson2009,Gullasch2011,Inci2015,Guelmezoglu2015,Gulmezoglu2016cross,Inci2016}, to break the integrity of secure enclaves~\cite{Gotzfried2017,Brasser2017sgx,Moghimi2017,Schwarz2017MGX,Dall2018cachequote}, monitor user interaction and keystrokes~\cite{Lipp2016,Schwarz2018KeyDrown,Wang2019Unveiling}, and build stealthy and fast covert channels~\cite{Xu2011,Lipp2016,Maurice2017Hello,Saileshwar2021Streamline}.
Two main techniques on cache attacks evolved with \PrimeProbe\cite{Osvik2006} and \FlushReload~\cite{Yarom2014Flush}.
Osvik~\etal\cite{Osvik2006} developed \PrimeProbe, where the attacker occupies cache lines and observes cache usage based on the cache eviction of the victim application.
\FlushReload requires a shared memory between attacker and victim, such as shared libraries~\cite{Yarom2014Flush}.
However, as \FlushReload works on the attacker's own addresses pointing to the same physical shared memory, \ie the attacker does not need to know the victim's address space layout or ASLR offsets, as binary offsets are used instead.

Further cache attacks were developed in JavaScript to spy on keystrokes, break memory randomization, and leak arbitrary memory in combination with transient-execution attacks~\cite{Oren2015,Lipp2016,Gras2017aslr,Kocher2019,VanSchaik2019RIDL,Roettger2020ridl,Tsuro2021SpectreJS,Agarwal2022Spookjs,Schwarzl2021Dynamic}.
Most of these attacks focus on hardware caches, \ie a \SI{64}{\byte} cache line granularity when attacking data and instruction caches.
For attacks on TLB caches, the spatial granularity can be \SI{4}{\kilo\byte}, \SI{2}{\mega\byte}, \SI{1}{\giga\byte}, or \SI{512}{\giga\byte}~\cite{Gruss2016Prefetch,Koschel2020,VanSchaik2018malicious,Lipp2022amd}.

In particular, for SGX, so-called controlled channels have been demonstrated as powerful attack primitives~\cite{Xu2015controlled,Vanbulck2017PTE,Moghimi2017} with high spatial and temporal resolution, as well as a very high accuracy.
Controlled channels are side channels running with elevated privileges, \eg kernel privileges, with a typical attack target being secure enclaves, that are protected against regular kernel access.

\subsection{Software caches}
On the software level, there are also caches that buffer data from slower memory components in faster memory components.
Software caches are used to cache frequent data like web requests in web servers~\cite{nginx,Redis2013MemtierBenchmark}, in-memory databases~\cite{memcached_website}, or keep frequently used database records~\cite{InnoDB_PhysicalStructure}.
These caches are limited in their size and follow similar replacement strategies to replace old data, such as LRU eviction~\cite{memcached_website}.
Caches like Nginx's request pool and Memcached~\cite{memcached_website} were exploited in memory-deduplication attacks~\cite{Bosman2016,Razavi2016,Schwarzl2022Remote}.

Operating systems use a software-level page cache to speed up accesses on disk-backed data by keeping the data buffered in main memory (DRAM)~\cite{Gorman2004,Gruss2019page}.
If the page cache is full, a page-replacement algorithm is used to replace pages and swap them out to disk~\cite{Bruno2013technet,Jiang2005}.
Linux and Windows offer functions to verify whether a specific virtual address is resident in memory or not, namely \texttt{mincore} and \texttt{QueryWorkingSetEx,Shared,ShareCount} respectively.
Gruss~\etal\cite{Gruss2019page} showed that cache attacks can be applied to the page cache by using either these functions or by measuring timing differences.
Consequently, the Linux kernel developers hardened the \texttt{mincore} implementation to only return page cache status information if the calling user can write to the underlying file, mitigating cache attacks using the \texttt{mincore} syscall.
The Windows kernel developers further restricted lower privileged processes to directly obtain information about the working set form higher privileged processes~\cite{Gruss2019page}.

On Linux, however, the system call \texttt{preadv2} can still be used to mount cache attacks~\cite{Corbet2019Fixing} in the same way as with the \texttt{mincore} syscall:
Using the \texttt{RWF\_NOWAIT} flag, an attacker can observe whether a page is resident in the page cache or not, yielding the same side-channel information as \texttt{mincore}.
The results of the \texttt{preadv2} templating attacks can be found in~\cref{sec:evaluation}.

\subsection{Automated Discovery of Side Channel Attacks}
Templating attacks have been first shown and mentioned on cryptographic primitives running on physical devices~\cite{Chari2002template,Rechberger2004Practical,Medwed2008Template}.
Brumley and Haka~\cite{Brumley2009} first described templating attacks on caches.
Doychev~\etal\cite{Doychev2013CacheAudit} presented a static analyzer that detects cache side-channel leakage in applications.
Gruss~\etal\cite{Gruss2015Template} showed that the usage of certain cache lines can be observed to mount powerful non-cryptographic attacks, namely on keystrokes.
Lipp~\etal\cite{Lipp2016} showed cache attacks and cache template attacks on ARM.
Van Cleemput~\etal~\cite{Van2017Adaptive} proposed using information gathered in the templating phase to detect and mitigate side channels.
Wang used symbolic execution and constraint solvers to speed up cache templating of cryptographic software ~\cite{Wang2017Cached}.
Schwarz~\etal\cite{Schwarz2018jstemplate} demonstrated template attacks on JavaScript to enable host fingerprinting in browsers.
Weiser~\etal\cite{Weiser2018} and Wichelmann~\etal\cite{wichelmann2018microwalk} showed that Intel PIN tools can be used to automatically detect secret-dependent behavior in applications, especially in a cryptographic context.
Wang~\etal\cite{Wang2019Unveiling} presented a similar automated approach to detect keystrokes in graphics libraries.
Carre~\etal\cite{Carre2019End-to-end} mounted an automated approach for cache attacks driven by machine learning.
With that approach, they were able to attack the \texttt{secp256k11} OpenSSL ECDSA implementation and extract \SIx{256} bits of the secret key.
Brotzmann~\etal\cite{Brotzman2019casym} presented a symbolic execution framework to detect secret-dependent operations in cryptographic algorithms and database queries.
Li~\etal\cite{Li2020SCNET} demonstrated a neural network to perform power analysis attacks automatically.
Yuan~\etal\cite{Yuan2021Automated} demonstrated that manifold learning can be used to detect and locate side-channel leakage in media software.

\subsection{Compiler-introduced Side Channels}
While developers typically focus on the source code level and care is taken to not introduce side channels there, the compiler translates the source code to a binary, essentially a different language.
However, this step can introduce program behavior that is not visible on the source level and introduces or amplifies side-channel leakage.
Page~\cite{Page2006A} demonstrated that dynamic compilation in Java leads to power side-channel leakage in a side-channel-secured library.
Simon~\etal\cite{Simon2018What} showed that mainstream C compilers optimizations can break cryptographically secure code by introducing timing side channels.
Brennan~\etal\cite{Brennan2020JIT} showed that timing side channels can be introduced by exploiting JIT compilation.

Due to this significant influence of compilers on side-channel leakage in binaries, they are also frequently used for new mitigation proposals against side-channel leakage~\cite{Coppens2009,Rane2015Raccoon,Pereida2016Make,Cauligi2017Fact,Garcia2017constant,Cauligi2020towards,Borrello2021Constantine}.

\section{\FullAttackName}\label{sec:attack}
For non-cryptographic applications, templating approaches~\cite{Gruss2015Template,Lipp2016,Wang2019Unveiling} can be used to find real-world exploitable leakage.
However, for large binaries like the Chrome browser with multiple shared libraries (\SI{220}{\mega\byte}), templating with fine granularity, \eg a cache line, becomes impractical.
Therefore, we present \TechniqueName, a layered approach taking advantage of the coarser granularity of some side channels, which is typically considered a disadvantage for the attacker.
In this section, we present the high-level view on \TechniqueName and show how \TechniqueName reduces the templating runtime by three orders of magnitude (\ie \SIx{1848}x).
We discuss implementation details and design decisions for our proof of concept implementation.

\begin{figure}[t]
  \centering
  \maxsizebox{\hsize}{!}{
    \input{images/attack_overview.tikz}
  }
  \caption{Overview of the \TechniqueName}
  \label{fig:attack_overview}
\end{figure}
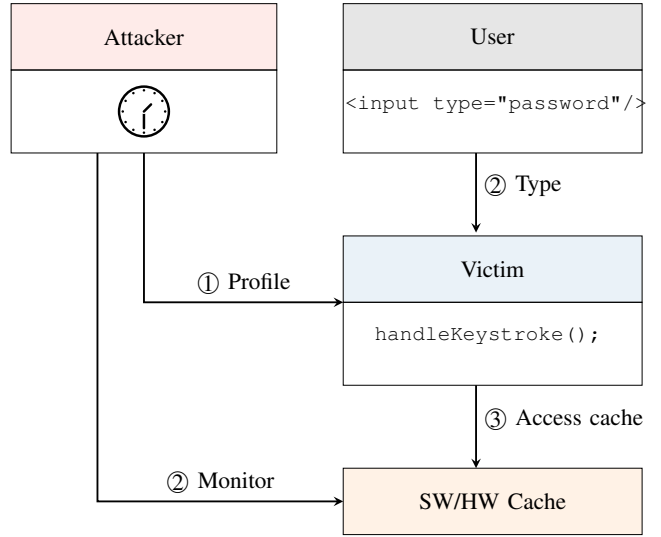

\subsection{Threat Model}\label{sec:threat_model}
Our threat model follows the standard template attack threat model~\cite{Gruss2015Template}, that distinguishes between the templating and the exploitation phase.

\paragrabf{Threat Model in the Templating Phase.}
For the \textbf{templating}, also known as the \textbf{profiling phase}, we assume the attacker was able to obtain a reasonably similar system, \ie with the same side channels as the victim system, such as the page cache and CPU cache.
The attacker has full control over this system and can template the same binary version, \eg obtained from package repositories or the vendor website, that also runs on the victim system.
The attacker can create templates for different binary versions and applications to have a suitable template in the exploitation phase subsequently.
In this phase, the attacker can also use any privileged or unprivileged mechanisms for the templating or improve its detection accuracy.

\paragrabf{Threat Model in the Exploitation Phase.}
In the \textbf{exploitation phase}, the attacker runs an \textbf{unprivileged} attack program on the victim's system, possibly under a separate user account.
Hence, we assume the victim application is started independently by the victim user, and cannot be started, stopped, or debugged by the attacker.
This also excludes ``preloading'', which, \eg on Wayland (the default Ubuntu display server), would allow monitoring all inputs to the application~\cite{Github2022wayland}.
For non-Wayland systems, we assume that the attacker cannot use other keylogging techniques (\eg on X11~\cite{Korpi2022xkbcat}), or Windows (\eg using the \texttt{getasynckeystate} API call~\cite{Wajahat2019Novel}), \eg due to system hardening or enforced security policies.
For our \textbf{keylogging scenario}, we assume that the victim user performs keystrokes manually and does not use scripted input generation, \eg auto-fillers or password managers.

\subsection{High-Level Attack Overview}
\Cref{fig:attack_overview} illustrates the steps of \TechniqueName.
First, the attacker templates the library and creates templates for the cache usage of different keystrokes (step 1).
After the templating phase, the attacker starts monitoring the software or hardware cache usage.
Depending on the cache activity, the attacker can infer if input was given or, in the optimal case, distinguish the keystrokes.

\subsection{Exploiting Different Spatial Granularity}
One novel aspect of \TechniqueName is to use the different spatial granularity of different side channels to our advantage.
The main focus here is on the templating phase, \ie the threat model for the templating phase applies.
In the following, we discuss how we extend cache templating from cache-line granularity, as used in previous work~\cite{Gruss2015Template}, over \SI{4}{\kilo\byte} page granularity~\cite{Gruss2019page} and \SI{2}{\mega\byte} page granularity, to a practical and generic multi-layered approach.
Our technique also extends to dimensions of \SI{1}{\giga\byte} steps and beyond but we are not aware of real-world victim binaries of such large size.

\paragrabf{64B Granularity.}
Previous cache template attacks~\cite{Gruss2015Template} used cache-line granularity (\SI{64}{\byte}).
One disadvantage of this approach is the runtime of the templating phase.
When templating a single cache line with \FlushReload, we observe an average runtime of \SIx{490} cycles ($n=1000000,\sigma_{\overline{x}}=20.35\%$).
On a \SI{4.0}{\giga\hertz} CPU, this would take \SI{122.5}{\nano\second}.
The Google Chrome binary has a file size of about \SI{210}{\mega\byte} leading to \SIx{2949120} addresses to template with \FlushReload.
This leads to a runtime of \SI{0.36}{\second} for templating every cache line once.
However, Gruss~\etal\cite{Gruss2015Template} describe that multiple rounds of \FlushReload are required to get reliable cache templating results.
Running an Intel 6700k CPU at \SI{4.0}{\giga\hertz} with a Ubuntu 20.04, templating \SI{1}{\mega\byte} of the Chrome browser (version 100.0.4896.60) with the provided implementation of Gruss~\etal\cite{Gruss2015Template}, we observe a runtime of \SI{817.652}{\second} for \SI{1}{\mega\byte} and a total runtime of \SIx{1.98} days for the full binary, including shared libraries, of \SI{210}{\mega\byte}.
Moreover, this templating tool only reports whether a certain address was cached or not and does not match the cache hits with the entered keystrokes.
To template, for instance, the \SIx{57} different common keys sequentially with the method by Gruss~\etal\cite{Gruss2015Template}, we would need an \textbf{impractical} total runtime of \SIx{113.17} days to obtain useful templates.
We conclude that such an approach is not feasible for browser developers as the code base changes frequently, and releases sometimes occur on a monthly basis~\cite{Chromium2022Speeding}.

\paragrabf{4kB Granularity.}
Page cache attacks exploit the operating system page cache, which works at a coarser granularity of \SI{4}{\kilo\byte}~\cite{Gruss2019page}.
Page cache attacks have the advantage of working independent of the underlying hardware.
To identify the exact memory locations causing leakage, they also resorted to templating.
However, they did not combine this information with timing differences from hardware caches.

Our intuitive idea here is to combine the \SI{4}{\kilo\byte}-granularity side channel with the more fine-grained side channel into a two-layered approach.
Hence, we \textbf{do not} template all cache lines on a \SI{64}{\byte} granularity but instead pre-filter memory locations on a \SI{4}{\kilo\byte} granularity.
Instead of \SI{2949120} memory locations, we then only need to monitor \SI{46080} memory locations for the Chrome example, \ie a templating runtime speedup of at least \SIx{64}.
In addition, the templating phase on the \SI{4}{\kilo\byte} granularity level implicitly identifies locations exploitable via the page cache.

In the templating phase, the attacker is not restricted as the code runs on the attackers own machine (\cf~\cref{sec:threat_model}).
Similarly to previous work, we can use the page cache side channel or privileged channels, \eg controlled-channel attacks~\cite{Xu2015controlled} via page-table bits~\cite{Vanbulck2017PTE}.
We used the idle bit as page tracker for Google Chrome (\cf~\cref{sec:implementation}) to be particularly useful as we have only a runtime of \SI{1.47} hours for all 57 key strokes.

\paragrabf{2MB Granularity.}
While the two-layered approach already brings a significant runtime speedup, our approach is more generic and extends to further layers.
Again, in the templating phase, the threat model (\cf~\cref{sec:threat_model}) permits the use of privileged channels, \eg controlled-channel attacks, but this time on the page directory layer.
Each page-table layer provides \texttt{referenced} bits that are set by the hardware when a location in this region is accessed.
This \SI{2}{\mega\byte}-granularity side channel is also exposed via various side channels~\cite{Gruss2016Prefetch,Koschel2020,VanSchaik2018malicious,Lipp2022amd}.
However, the activity on \SI{2}{\mega\byte} pages can be observed via the Page Middle Directory paging structure and the \texttt{referenced} bit.
We use the tool PTEditor~\cite{SchwarzPteditor} to clear the \texttt{referenced} bit in the PMD.
Checking and clearing the referenced bit of the PMD, \ie a \SI{2}{\mega\byte} page, leads to a runtime of \SI{661.965}{\nano\second} ($n=1000000,\sigma_{\overline{x}}=0.049\%$).
Similar to the templating with \SI{4}{\kilo\byte} granularity, we assume that multiple repetitions are required to observe stable results and remove false positives.
Hence, to template the \SIx{57} different common keys in Chrome with \SIx{20} repetitions per key, we estimate the total templating to be about \SIx{0.15} seconds.

\subsection{Beyond Huge Pages}
The concept of \TechniqueName extends to arbitrarily coarser granularities as long as a channel exists.

\paragrabf{\SI{1}{\giga\byte} Granularity.}
For the \SI{1}{\giga\byte} granularity level, we experimentally validated that we can again use controlled-channel attacks~\cite{Xu2015controlled,Vanbulck2017PTE}, albeit this time on the page-directory-pointer-table level.
On this level, one entry (and thus one \texttt{referenced} bit) covers a \SI{1}{\giga\byte} region.
Following a similar approach as for the previous levels, we use PTEditor to template and clear the referenced bit, for the single offset in a \SI{1}{\giga\byte} range.
The runtime for checking and clearing the referenced bit in the PUD is the same as for the PMD (\SI{661.965}{\nano\second} ($n=1000000,\sigma_{\overline{x}}=0.049\%$)).
Note that this layer of \TechniqueName may become relevant in the future with constantly growing binaries and libraries.\footnote{The Google Chrome binary had \SI{100}{\mega\byte} in 2017 and \SI{180}{\mega\byte} in 2022, showing an increase of \SI{80}{\percent}.}

\paragrabf{\SI{512}{\giga\byte} and \SI{256}{\tera\byte} Granularity.}
The intuitive extension of our approach also uses the \texttt{referenced} bits on higher page-table layers, \ie the page-map level 4 to template in \SI{512}{\giga\byte} steps, and the page-map level 5 (if available) to template in \SI{256}{\tera\byte} steps.
We verified that we can indeed observe activity in this range via controlled-channel attacks~\cite{Xu2015controlled,Vanbulck2017PTE}, \ie via the \texttt{referenced} bits, showing that \TechniqueName can be extended to a 5-layer templating phase.
The runtime for checking and clearing the referenced bit in the PML4 is the same as for the PUD (\SI{661.965}{\nano\second} ($n=1000000,\sigma_{\overline{x}}=0.049\%$)).

We emphasize that scanning layers that exceed the binary size, \eg the \SI{1}{\giga\byte} layer for a \SI{180}{\mega\byte} binary, provides no additional information and does not reduce the search space, as the search will always proceed to the next smaller layer for the entire memory range then.
Therefore, in the evaluation, we skip all layers that exceed the binary size.

\subsection{Templating Phase Implementation}\label{sec:implementation}
The high-level idea is that the templater tracks the usage of pages and then actively filters out pages that are not related to keystrokes to reduce the search space of pages to template and, as a result, reduce the overall runtime of the templating.
We implement our templater in Python, and provide it as open source in our Github repository~\cite{Anonymous2022blind}.
The templater takes as input the set of different keys, the PID or process names that should be monitored, and the number of samples per key.

\Cref{alg:templater} summarizes the steps of the templating.
First, the templater runs a warmup phase, where all keystrokes to template are entered once to load all related memory locations into RAM.
Then the templater collects all the memory mapping information from all files from the target processes where activity has been found.
These memory mappings include all shared libraries.
The templater generates random key sequences based on the set of keys to template.
For each key in the sequence, the templater iterates over all the memory locations on the current granularity level, and resets the access information, \ie resets the \texttt{referenced} or \texttt{idle} bit, or flushes the cache line depending on the side channel used.
Based on the number of samples performed, the templater computes the hit ratio for each location.
Subsequently, the templater repeats this step for all memory locations above a specific hit ratio with the next lower spatial granularity.
With this search strategy, the templater continues down to the lowest level, where only regions are templated that showed activity on coarser granularities.
On the lowest level, the templater then obtains a hit ratio for each single cache line.

\begin{algorithm}[!h]
  \caption{\TechniqueName Templating Algorithm}
  \label{alg:templater}
  \renewcommand{\algorithmicrequire}{\textbf{Input:}}
  \renewcommand{\algorithmicensure}{\textbf{Output:}}
  \begin{algorithmic}[1]
    \REQUIRE Set of keys $K$, target PIDs $P_n$, number of samples $N$  %
    \ENSURE hit ratio matrix of all memory mappings $H$    %

    \STATE Enter all keys in $K$ once // Warmup
    \STATE Collect all valid memory mappings of $P_n$ \\(possibly from previous layer)

    \FOR{$i=0; i < N;i++$}
    \FOR{each $k \in K$}
    \STATE Reset memory mappings (reset referenced/idle bits or flush)
    \STATE Enter key $k$
    \STATE Check state for all present memory mappings (via interface or timing)
    \STATE Compute hit ratios for $k$ and update $H_k$
    \ENDFOR
    \ENDFOR
    \RETURN $H$, and repeat algorithm for next layer
  \end{algorithmic}
\end{algorithm}

\paragrabf{Linux.}
On the upper layers, we start by obtaining the memory mappings for the target process.
On Linux, we read these mappings from procfs (with root privileges in line with the threat model).
We group the memory locations then according to the most coarse granularity we use in our templating.
By using the \texttt{referenced}-bit side channel according to \cref{alg:templater}, we narrow down the set of memory locations for the next layer.

\paragrabf{Windows.}
On Windows, we can also obtain a list of memory mappings using the \texttt{EnumProcessModules} PSAPI call, which lists all loaded libraries and executable programs.
To retrieve the actual sizes of the libraries and executables, the \texttt{GetModuleInformation} PSAPI call is used.
Subsequently, we follow the same process using the \texttt{referenced}-bit side channel as on Linux to narrow down the set of memory locations using \cref{alg:templater}.
Subsequently, we continue with the next layer.

\subsubsection{\SI{4}{\kilo\byte} Page Granularity}
While for the upper layers, we used PTEditor~\cite{SchwarzPteditor} to read the referenced bits, we implemented a more optimized approach for the page granularity.

\paragrabf{Linux.}
We implement a page usage tracker which iterates over all active mappings, reads the idle bit for the corresponding physical page from \texttt{/sys/kernel/mm/page\_idle/bitmap} and checks if the page was accessed.
We start by resetting the bit so that the page usage tracker is ready.
We use the Python3 \texttt{keyboard} library to inject keystrokes into an input field.
After the templater performs the sequence of keystrokes, we check all pages that are still in the candidate list for activity.
If we now observe a \SIx{1} at the page offset in the bitmap, the page was not accessed.
Conversely, if we observe a \SIx{0} at the page offset, we reset the page offset, and add it to the set of correlated pages to track on the next layer.
Note that this approach is fully hardware-agnostic as this feature is implemented in software in the Linux kernel.
After each iteration, we reset the state again by marking the pages as idle again and repeat the measurements.

In case of a sequential read access pattern, the Linux kernel speculatively prefetches further pages of the same file after a new page was added to page cache.
This optimization is called \texttt{read\-ahead}~\cite{HaloLinux2022Page}.
On Ubuntu 20.04 (kernel 5.4.0), the default read-ahead size is \SI{128}{\kilo\byte} and can be found in the sysfs (\texttt{/sys/block/\\<block\_device>/bdi/read\_ahead\_kb}).
For file-mappings the kernel performs a different optimization called \texttt{read-around}.\footnote{https://elixir.bootlin.com/linux/v5.4/source/mm/filemap.c\#L2437}
There, the kernel prefetches pages surrounded by the page causing the pagefault \eg \SIx{16} pages before the page causing the pagefault and \SIx{15} pages after.
To reduce triggering read-ahead prefetching for sequential reads, we use the madvise system call with the \texttt{MADV\_RANDOM} flag to indicate a random read order.

Overlapping event (\ie keystroke) groups for the current candidate page and pages that might trigger the readahead of the current candidate page could cause false positives in the Linux case.
In addition, if the number of readahead suppress pages is too small, false positives can occur.
Our classifier tries to reduce the number of false positives by checking out the read-ahead/read-around windows and systematically rule out other keystrokes.
Based on the results of the templater, the classifier actively accesses surrounding pages from the target page to suppress the read-ahead/read-around optimization.
Note that the read-ahead and read-around windows might overlap for some keys.
If the keys to template are not on the same \SI{4}{\kilo\byte}-page, we can still distinguish two keys by checking the first and last surrounded pages being accessed.
The templater actively creates warnings in the templating phase in case the keys are still indistinguishable.

\paragrabf{Windows.}
Windows uses a different page replacement strategy with global and local working sets~\cite{Russinovich2012}.
To build the page usage tracker on Windows, we use the PSAPI call \texttt{QueryWorkingSetEx} and monitor the \texttt{Shared},\texttt{ShareCount} and \texttt{Valid} flags.
If the page is marked as valid and shared and the share count is larger than \SIx{1}, we mark the page as used.
To perform the reset, the \texttt{EmptyWorkingSet} PSAPI call is used to remove the pages from all workings sets.
This PSAPI call is only available for unprotected processes, which is no issue during the templating phase (\cf~\cref{sec:threat_model}).

On Windows, we observed no prefetching optimization within working sets.
Therefore, read-ahead is no problem for measuring the hit ratio and detecting the exact offset on this level.
Alternatively, the templating could also be performed via controlled side channels~\cite{Xu2015controlled,fu2017lapd,Shih2017tsgx}, tracing tools such as Intel PIN, machine learning~\cite{Wang2019Unveiling,Carre2019End-to-end} or architecturally monitoring the accesses of pages using PTEditor~\cite{SchwarzPteditor}.

\paragrabf{Classifier.}
We further try to restrict the search space further on the \SI{4}{\kilo\byte} level.
On the \SI{4}{\kilo\byte} level, we collect the page-hit ratios for all events (\ie keys) and pages, showing the link between event and observable page hit.
The templater also templates a dummy idle event that sleeps for 30 seconds to measure which hit ratios are above the system's noise floor.
This idle event will not be linked to any page hit but rather should represent unrelated noise that the templater might have missed during the sampling of the other events.
Our classifier links events or groups of events with single page hits to keep the number of observed pages as low as possible.
This is a trade-off between search time and completeness of the search that can be chosen differently for any \TechniqueName on any target application.
Furthermore, a more sophisticated attack could increase detection accuracy from monitoring multiple pages or cache lines for each event.
However, we decided to use the search-time-optimized path, as side-channel attacks typically cannot observe an arbitrary amount of memory addresses anyway, \ie we focus on a more practical set of leaking addresses.

The algorithm to find a suitable page hit describing an event $e$ works as follows:
\begin{enumerate}
  \item We take the vector of all page-hit ratios related to event $e$ and subtract the sum of the page-hit-ratio vectors of all other events (noise) from it.
        The resulting vector describes how strongly accesses to a certain page (index in the vector) are correlated with an event $e$.
  \item We select the page with the highest score as a candidate.
        \begin{enumerate}
          \item Should the resulting candidate fail to fulfill our minimum score requirements (\ie not above the \textbf{location-specific} noise floor), our algorithm merges events to find a suitable page to recognize this group of events.
          \item We treat the target event and the $M - 1$ remaining events (exluding the idle event) with the highest page-hit ratios per page as one group.
                The difference is that we now have an event group $E = \{e_1, ..., e_M\}$ instead of a single event.
                The group's score is the minimum score of its events.
                After merging, we use the same classification scheme as in step 1.
          \item Should we still fail to find a suitable page, we return to step 2.1, \ie we merge further events until we succeed, or we end the classification for the event.
        \end{enumerate}
  \item While running, the classifier also collects information on potential read-around prefetching pages to filter them out.
\end{enumerate}
After a successful classification, the attacker has a mapping of pages to events (\ie key) and groups of events (\ie groups of keys).
Subsequently, we can use this information and the trace from the side-channel attack to build our keylogger.

\section{Compiler- and Linker-introduced Spatial Distance in Binaries}\label{sec:compiler_sc}
Before we evaluate \TechniqueName, we present one significant leakage-facilitating effect that we discovered while applying \TechniqueName on a variety of targets.
This effect is particularly critical as it originates in compiler optimizations in LLVM/clang that are enabled by default and cannot be disabled via compiler flags.
Compiler optimizations aim for a minimal program runtime, small memory footprint, and small binary size.
Moreover, linker optimizations try to further optimize the binary in the linking stage.
This is especially useful for large software projects such as browsers.
We primarily found two effects to facilitate cache side-channel leakage:
One is \textbf{data deduplication during compilation and linking}, the other is the first-come-first-serve data placement introduced by the \textbf{dead-stripping} mechanism.
While memory deduplication at runtime has been explored as a security risk already (\cf~\cref{sec:dedup}), data deduplication (\eg of strings) during compilation is not widely known and its security implications are entirely unexplored.
The security of constant-time implementations has been analyzed for side channels being introduced by compilers~\cite{Page2006A,Van2017Adaptive,Simon2018What,Brennan2020JIT}.
In this section, we first show that dead-stripping (\cf~\cref{sec:deadstripping}) can facilitate and amplify side channels on read-only strings used in C/C++ programs.
We then show that deduplication during compilation (\cf~\cref{sec:dedupcompilation}) and linking (\cf~\cref{sec:deduplinking}) can amplify this effect by increasing the chance that secret-dependently accessed victim data is placed in an attacker-facilitating way.
In addition to compiler optimizations, deduplication can als be performed at the linking stage.
The spatial distance between secret-dependent accesses can be introduce by both compiler and linker optimizations.
We present two scenarios that we also found in widely used real-world applications, where the placement of read-only data, especially strings, amplifies side-channel leakage dramatically.

\subsection{Dead-Stripping}\label{sec:deadstripping}%
\begin{minipage}[c]{0.95\columnwidth}
\begin{lstlisting}[language=C,style=customc,label={lst:access},caption={Compiler populates .rodata based on accesses in code}]
struct MapEntry {
  const char* key;
  const char* value;
};
#define LANGUAGE_CODE(key,value) \
  { key, value }
#define MAP_DECL constexpr MapEntry mappings[] 
=  MAP_DECL {
    LANGUAGE_CODE("KeyA","DataA"),
    LANGUAGE_CODE("KeyB","DataB")
  };
#undef MAP_DECL
void string_func(vector<string>& vec) {
  //Key A
  MapEntry k1 = mappings[0];   
  vec.push_back(k1.value);
  // Padding string
  string padding = "<64-byte-string>";   
  vec.push_back(padding);
  //Key B
  MapEntry k3 = mappings[1];   
  vec.push_back(k3.value);
}
\end{lstlisting}
\end{minipage}

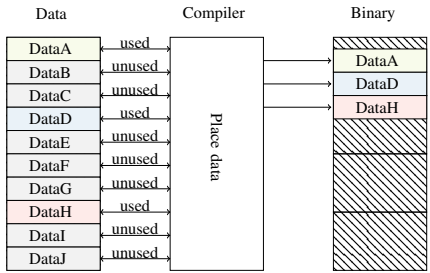
\begin{figure}[t]
  \centering
  \maxsizebox{0.8\hsize}{!}{
    \input{images/dead_stripping.tikz}
  }
  \caption{Dead stripping in combination with first-come-first-serve population of the .rodata section in the binary.}
  \label{fig:dead_stripping}
\end{figure}

Lookup tables are frequently used in applications to speed-up memory accesses and store constant data like locality strings.
For the developer, it is not transparent how constants are stored in the compiled binary.
Thus, even if the code seems to hide a cache-side channel on \SI{64}{\byte} granularity, the compiler might reorder strings and add more spatial granularity between items of the same lookup table.
One optimization to reduce the binary size is to only populate the read-only data section if the compiler observes that only certain indices of a lookup table are accessed.
\Cref{fig:dead_stripping} illustrates how data can be re-ordered in the use case of dead-stripping.
If the developer uses a macro to dynamically populate a lookup table, \eg with key mappings or similar, compilers do not insert all elements into the read-only section of the binary to reduce the binary size.
\Cref{lst:access} lists an example of code that seems to have co-located strings mappings array in the .rodata section.
However, as the compiler is free to populate the readonly section in the binary, it is possible that the compiler introduces leakage here that is not visible on the source level.
While without this optimization, an attacker could not distinguish whether \texttt{KeyA} or \texttt{KeyB} was accessed with \FlushReload in the original code snippet. The compiler creates a spatial distance exposing this secret information to the attacker.
We evaluate our test program on Clang 10 and GCC 9.4.0 with the optimization levels \texttt{O0-O3,Os}.
For Clang, we observe that on every optimization level and also without optimizations, the padding string is between the DataA and DataB strings, \ie \verb|DataA.<64-byte-string>.DataB.|.
The \verb|.| indicates a NULL-byte in the memory layout for C strings.
Clang compiler traverses all functions and populates the .rodata depending on the fixed indices for the array lookup.
GCC with the O0/O1 flags populates the .rodata directly with all strings from the mappings array, \ie \verb|DataA.DataB.<64-byte-string>|.
Therefore, an attacker would not be able to observe which cache line was accessed.
For the other optimization levels in that concrete example, the small strings \verb|DataA|,\verb|DataB| would be directly stored as immediate in the binary, \eg \texttt{"mov    DWORD PTR [rsp+0x50], 0x756c6156"} and then the full string on the heap is allocated.

\subsection{Data Deduplication during Compilation}\label{sec:dedupcompilation}
\begin{figure}[t]
  \centering
  \maxsizebox{0.8\hsize}{!}{
    \input{images/compiler_string_merging.tikz}
  }
  \caption{String deduplication in the compiler causing spatial distance in .rodata section of the binary.}
  \label{fig:compiler_merging}
\end{figure}
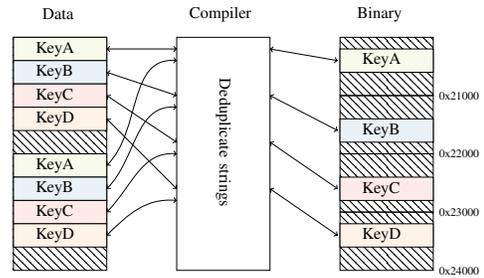

Another optimization facilitating cache attacks, also in combination with the dead-stripping we just discussed, is data deduplication during compilation.
Deduplicating strings can reduce the binary size significantly but also the memory resident size when running the program, as strings do not have to be kept in memory multiple times.
\Cref{fig:compiler_merging} demonstrates how string deduplication can introduce spatial distance in sections of the binary, for instance, the .rodata section.
C/C++ compilers deduplicate strings that occur more than once in the source code.
\Cref{lst:dedup} illustrates a situation where the string deduplication optimization can be used.
Both the lookup table \texttt{mappings} and the function \texttt{string\_funcA} contain the string \texttt{DataA}.
Since the compiler again traverses over the functions and \texttt{DataB} is first inserted into the .rodata section.
Again a constant padding string could cause \texttt{DataA} and \texttt{DataB} to be located in different cache lines.
Before the compiler inserts \texttt{DataB} (\texttt{mappings[1]} in \texttt{string\_funcB}), the compiler checks for duplicates and only points to the existing occurence of \texttt{DataB} in the .rodata for all future usages.
Again we evaluate this code constellation for GCC and Clang.
For Clang, we observe again for all optimizations levels the ordering \verb|DataA.<64-byte-string>.DataB| in the .rodata section.
For GCC, we observe the same result that for optimization levels O0/O1, both values are populated next to each other in the .rodata (\verb|DataA.DataB|).
Again for the higher optimization levels, the small strings are encoded as immediates.

\begin{minipage}[c]{0.95\columnwidth}
\begin{lstlisting}[language=C,style=customc,label={lst:dedup},caption={Strings are deduplicated in the binary and could lead to spatial distance between readonly-strings in the same array.}]
struct MapEntry {
  const char* key;
  const char* value;
};
#define LANGUAGE_CODE(key,value) \
  { key, value }
#define MAP_DECL constexpr MapEntry mappings[] 
= MAP_DECL {
  LANGUAGE_CODE("KeyA","DataA"),
  LANGUAGE_CODE("KeyB","DataB")
};
#undef MAP_DECL
void string_funcA(vector<string>& vec) {
  string local_ro_string = "DataB"; 
  vec.push_back(local_ro_string);
  string padding_string = "<64-byte-string>"; 
  vec.push_back(padding_string);
}
void string_funcB(vector<string>& vec) {
  //KeyA
  MapEntry k1 = mappings[0]; 
  vec.push_back(k1.value);
  //KeyB
  MapEntry k2 = mappings[1]; 
  vec.push_back(k1.value);
}
\end{lstlisting}
\end{minipage}

\subsection{Deduplication in the linking step.}\label{sec:deduplinking}
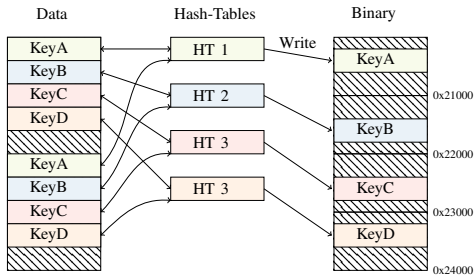
\begin{figure}[t]
  \centering
  \maxsizebox{0.8\hsize}{!}{
    \input{images/linker_string_merging.tikz}
  }
  \caption{String deduplication in the linker causing spatial distance in .rodata section of the binary.}
  \label{fig:linker_merging}
\end{figure}

As we showed, string deduplication can cause spatial distance between strings and enable side-channel attacks in the compile step.
For large software projects such as the Chromium project, it is important also to save time while linking together a large amount of object files.
However, for such large software projects, the runtime of the linker is crucial.
Linkers are also merging strings as an optimization in the linking time.
To perform fast lookups of strings in the linking step, hash tables are used~\cite{Ueyama2019lld}.
Since 2017, lld uses multiple hash tables instead of one large one to enable concurrency.
Concurrency in the linking step brings a speedup in the linking time.
However, as there are multiple tables, inserting merged strings can cause a different layout for strings in the .rodata section than in the final linked binary.
\Cref{fig:linker_merging} illustrates how the concurrent merging can lead spatial distance in the final binary.
In the highest optimizations of \texttt{lld} linker, \ie \verb|-O2|\cite{nxmnpg2022Manual}, the linker merges duplicate substrings contained in larger strings.
The smaller substring will be removed, and the tail of the larger string used to index the substring.
The security implications of string deduplication need to be considered in software projects since large spatial distance between secret dependent values such as different key inputs can lead to leakage of all user input, as we show in~\Cref{sec:evaluation}.

\section{Evaluation} \label{sec:evaluation}
In this section, we evaluate our templater on large binaries such as browsers that have not been targeted with templating attacks so far.
We also focus on widespread Chromium-based products and demonstrate that they are susceptible to \TechniqueName.
We analyze the root cause for the leakage and show that it is caused by a compiler optimization.
\Cref{tab:evaluated_programs} lists all the evaluated applications, including the Chromium-based browsers and applications, Firefox and LibreOffice Writer.

\paragrabf{Templating of HTML form input fields Google Chrome.}
We first run our templating tool while generating keystrokes.
We run our templater on an Intel i7-6700K with a fixed frequency of \SI{4}{\giga\hertz} running Ubuntu 20.04 (kernel 5.4.0-40) on Google Chrome version 100.0.4896.60.
To get more accurate results during the templating phase, we recommend dropping the active caches before the execution of the templater via procfs (\texttt{/proc/sys/vm/drop\_caches}).
Moreover, we blacklist file mappings from the \texttt{/usr/share/fonts/} as they lead to unstable results during the evaluation phase.
Our templater traces 57 different key codes of a common \texttt{US\_EN} keyboard in HTML password fields.
For each key code, we sample \SIx{20} times.
On average, we observe a runtime of \SIx{1.47} hours ($n=10,\sigma_{\overline{x}}=0.33\%$) for 57 key codes, including the time for key classification.
For a single character, the runtime is \SIx{92} seconds.
We use the \texttt{ldd} to find all shared libraries used in Google Chrome to determine the overall file size of libraries to template.
Including the main binary, the total size of memory mappings to scan is \SI{209.81}{\mega\byte}.
We estimate the runtime of the public implementation for cache template attacks~\cite{Gruss2015Template} to \SIx{113.17} days. %
Our templater brings a speedup in the templating runtime of factor \SIx{1484}.
Using the proposed optimizations of Gruss~\etal\cite{Gruss2015Template} leveraging parallelism in the templating phase, we can achieve a runtime of \SI{17}{\second} per megabyte.
Thus, the templating time could be improved to \SIx{2.35} days. %
However, this approach appears to be noisier and was not merged in the public implementation of cache templating attacks~\cite{Gruss2015Template}.

\paragrabf{Leakage Source in Chrome.}
As we discovered the page offsets related to the different keystrokes, we want to find the exact cache line causing the cache leakage.
We extend our monitor with \FlushReload to determine the cache line within the page.
To speed up the templating time and reduce the noise for cache lines, we disable most of the Intel prefetchers by writing the value \texttt{0xf} to MSR \texttt{0x1a4}~\cite{Intel_DisableHWPrefetcher}.
We map the Chrome binary as shared memory and perform \FlushReload on all mapped cache lines to determine the corresponding cache lines for each key.

\begin{figure}[t]
  \centering
  \maxsizebox{0.8\hsize}{!}{
    \input{images/chrome_leakage_rodata.tikz}
  }
  \caption{Illustration of strings in the .rodata section for key codes that cause the cache input leakage.}
  \label{fig:leakage_rodata}
\end{figure}
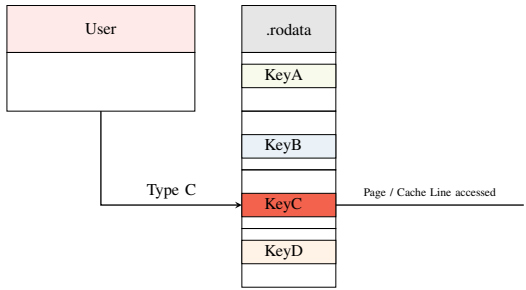

We analyze the Chrome binary and lookup the offsets causing the leakage for a specific key stroke.
Each cache line causing the leakage of a certain character contains a string for the key event, \eg ``KeyA''.
By reading Chrome's ELF header, we observe that all offsets lie in the read-only data (\texttt{.rodata} section of the binary.
The leakage source are key-dependent accesses to the key code strings in the dom code table,\footnote{\url{https://source.chromium.org/chromium/chromium/src/+/main:ui/events/keycodes/dom/dom_code_data.inc}} \eg \texttt{DOM\_CODE(0x070004, 0x001e, 0x0026, 0x001e, 0x0000, "KeyA", US\_A);}.
As these strings appear multiple times at different locations in the code, the location in the \texttt{.rodata} depends on the compiler.
\Cref{fig:leakage_rodata} illustrates the leakage source for a user typing in a certain character and the corresponding DOM\_CODE for the UI event.

To verify if the leakage is related to string deduplication, we download the Chromium source, disable the string deduplication \texttt{-fno-merge-all-constants} and rebuild the Chromium browser.
We still observe, that the single keystrokes are spread over multiple pages in the \texttt{.rodata} section.

As a next step, we analyze the compiled object files after the build process.
We observe that the created object file \texttt{keycode\_converter.o} still contains all the key event strings adjacent to each other in the binary.
This indicates that the linker introduces the spatial distance between key event strings.

We perform a binary search on older Chrome binaries from a public Github repository containing archived Chrome Debian packages~\cite{Github2022ChromeArchive} to see when the spatial distance for key event strings was introduced.
As a result, we observe that between version 63 and 64 of Chrome (year 2017), the single key event string was placed in the \texttt{.rodata} at different \SI{4}{\kilo\byte} pages.
According to~\cite{Ueyama2019lld}, the linker optimizations have been constanly improved since 2017.
As discussed in~\Cref{sec:compiler_sc}, the parallelism in string deduplication can also cause spatial distance between key events.
Disabling the string merging optimization of the linker with \verb|-Wl,-O0| for a current Chromium version removes the spatial distance between the key event strings.
Re-enabling the string merging optimization of the linker, \ie \verb|-Wl,-O1|, the spatial distance reappears as strings are again deduplicated.
This confirms that one of the effects we exploit is introduced by the linker.

In comparison to state-of-the-art keyloggers on Linux like xkbcat~\cite{Korpi2022xkbcat}, our keylogger does not rely on running as the same user within the same X-session.
We verify this by running our keylogger as a different user and can still recover the keys from the Chrome browser.
Note that more sophisticated cache attacks like \PrimeProbe can be performed to enable JavaScript-based attacks~\cite{Osvik2006,Vila2019theory,Agarwal2022Spookjs}.

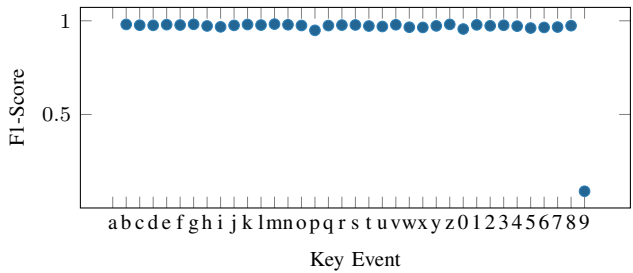
\begin{figure}[t]
  \centering
  \maxsizebox{\hsize}{!}{
    \input{images/fscore_fr.tikz}
  }
  \caption{F-Score per key using \FlushReload for all alphanumeric characters in Chrome.}
  \label{fig:fscore_fr}
\end{figure}

\paragrabf{Keylogging in Google Chrome with \FlushReload.}
We run our monitor in three experiments for 180 seconds with all lowercase alphanumeric characters and observe cache activity for every single keystroke.
The first experiment runs with fast user input with \SI{1}{\milli\second} between each keystroke.
We count cache hits following a keystroke as true positives if they occur on the cache line that is correct according to our template, and as false positive otherwise.
To obtain the number of false positives, we run the monitor in a second experiment without performing any keystrokes in the input field, \ie idling.
To complete our data on false negatives and true positives, we run the monitor in a third experiment while performing user input with \SI{1}{\second} between each keystroke.
Over the total 540 second measurement time frame, we observed no false negatives.
\Cref{fig:appendix:ratio_pc} (Appendix) shows the cache-hit ratio for the cache lines detecting lowercase letters in Chrome.
\Cref{fig:ratio_fr_digits} shows the cache-hit ratio for the cache lines detecting numeric digits in Chrome.
As shown from \Cref{fig:ratio_fr_digits}, the different digits can be highly-accurately classified.
As can be seen, digit \SIx{9} causes constant noise for all the different keys as in that version of Google Chrome.
The cache-line is constantly accessed also in an idle state.
The F-Score is the harmonic mean of precision and recall.
\Cref{fig:fscore_fr} illustrates the F-Score for all alphanumeric characters.
From all the \SIx{36} alphanumeric keys, this is the only character causing constant noise on Google Chrome.
Moreover, we observe that a single keystroke causes up to three cache hits.
These cache hits could be related to the window events \texttt{key\_up},\texttt{key\_pressed} and \texttt{key\_down}.
To avoid printing the same character multiple times, a cache miss counter between the keystrokes can be used~\cite{Gruss2015Template}.
Note that multiple cache lines can be considered to further increase the accuracy of the keylogger~\cite{Lipp2016,Wang2019Unveiling,Carre2019End-to-end}.

\begin{figure}[t]
  \centering
  \maxsizebox{\hsize}{!}{
    \input{images/ratio_digits_fr.tikz}
  }
  \caption{Cache-hit ratio using \FlushReload for all digits letters in Chrome.}
  \label{fig:ratio_fr_digits}
\end{figure}

\paragrabf{Keylogging with the page cache.}
To demonstrate that the Chrome leakage is not specific to a certain CPU, we run our keylogger on Chrome version 99.0.4844.84.
Our test device runs Ubuntu 20.04 (kernel 5.18.0-051800-generic), equipped with an AMD Ryzen 5 2600X CPU, \SI{16}{\giga\byte} of RAM, and a Samsung 970 EVO NVME SSD.
We circumvent the read-around and read-ahead optimization as explained in~\Cref{sec:implementation}.
The keylogger uses the keystroke template for the main Chrome binary and monitors the page cache utilization for the corresponding pages using the \texttt{preadv2} syscall.
It then reports the detected activity as keystrokes and subsequently evicts the page cache.
While the page cache attack using the \texttt{mincore} syscall was able to observe keystrokes on a fine temporal granularity, we observe that using \texttt{preadv2} comes with practical limitations.
In particular, with large eviction set sizes, guessed by the attacker, we conclude that only very slow keyboard interaction with gaps of \SI{2}{\second} and more can be observed.
However, our evaluation of the page-cache side channel is generic and would also apply to scenario where the \texttt{mincore} syscall is available, which allowed fast and non-destructive continuous probing.

\begin{figure}[t]
  \centering
  \maxsizebox{\hsize}{!}{
    \input{images/fscore_pc.tikz}
  }
  \caption{F-Score per key event using page cache attacks for all alphanumeric characters in Chrome.}
  \label{fig:fscore_pc}
\end{figure}
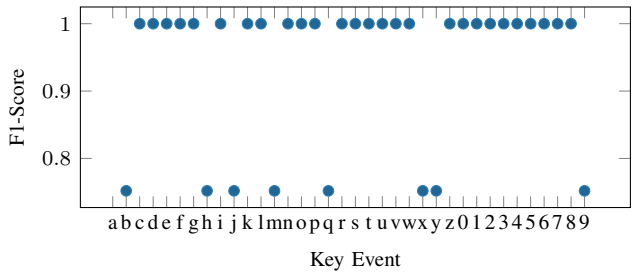

Based on the page cache accesses, we compute the page-hit ratio for Google Chrome over the page cache.
\Cref{fig:appendix:ratio_pc} (Appendix) shows the page-hit ratio for the page cache detecting alphanumeric letters in Chrome.
For the Chrome version, we observe that the characters \texttt{b,m,9,h,y,x} are grouped and cannot be uniquely distinguished.
We again perform the experiment in three phases to determine true positive, false positive, and false negative rate, by simulating fast, slow, and no user input.
\Cref{fig:fscore_pc} shows the F-Score for all alphanumeric characters running the page cache attack.
While most characters have very high F-Scores, the character group \texttt{b,m,9,h,y,x} has a lower F-Score due to false positives when other keys are pressed.
Also, same as in the \FlushReload attack, the character \texttt{9} suffers from a high number of false positives, negatively impacting the F-Score.

\paragrabf{Electron.}
As we observed the leakage of keystrokes within the Chrome binary, we further analyze Chromium-based applications like the Electron framework.
As Chromium-based applications use the same handling for keystrokes for all applications, we can directly scan the \texttt{.rodata} section for the keystroke mappings to find the offsets.
We evaluate the templates on Chromium, Threema, Passky, VS-Code, Mattermost, Discord and observe similar leakage rates to Google Chrome with F-Score of over \SI{85}{\percent}.
\Cref{tab:evaluated_programs} contains the high F-Scores for the different applications.
We run our monitor on the lowercase alphanumeric keys and observe a similar leakage rate with \FlushReload.
Based on these very clear results, we deduce that in principle all Electron applications are susceptible to \TechniqueName and full key recovery attacks.

\paragrabf{Chromium Embedded Framework.}
The Chromium Embedded Framework (CEF) is widely used and another interesting target for \TechniqueName.
While Electron directly uses the Chromium API, CEF tries to hide the details of the Chromium API~\cite{Electron2022ElectronInternals}.
CEF is actively run on more than 100 million devices~\cite{CEF2022Chrome}.
We target Spotify, and the Brackets editor application, which are both based on CEF.
To attack a CEF application, an attacker needs to read out the \texttt{.rodata} section from the shared library \texttt{libcef.so}.
We run our monitor again with \FlushReload and observe an F-Score of \SI{0.96}{\%} over the lowercase alphanumeric characters.
For Brackets (1.5.0) we observe, that the \texttt{libcef.so} was built with an older linker version as the different key-event related strings for the lowercase alphanumeric characters are co-located in three different cache lines.
Therefore, we consider all CEF applications to be susceptible to cache templating in principle.
We observe an F-Score of \SI{94}{\percent} for detecting key events.
However, we also observe that hardware prefetching practically thwarts the distinction of different blocks in this scenario more than in the other attack scenarios, leaving only inter-keystroke timing attacks as an option for the attack phase.

\paragrabf{Firefox.}
We also templated the Firefox binary with \TechniqueName and identified locations with cache activity upon keystrokes.
While we observe cache activity for each keystroke in the shared library \texttt{libxul.so} (offset: 0x332d000), we did not find leakage to distinguish keys reliably.
Firefox uses a completely different build system, \ie optimizations such as data deduplication and dead-stripping may still be applied but will not behave exactly the same as with LLVM/clang.
However, using the found binary offset, an attacker can determine whether a user is typing and perform an inter-keystroke timing attack~\cite{Song2001,Ristenpart2009,Zhang2009keystroke,Diao2016,Gruss2015Template} to recover the keystrokes.
We run the monitor again and probe the address that has the most hits within the page on different keystrokes.
The accuracy we observed for such an attack is \SI{96}{\percent}.

\paragrabf{LibreOffice Writer.}
We profile the LibrefOffice Writer version 6.4.2 on our Linux setup.
Our profiler shows that the library \texttt{libQt5XcbQpa.so.5.12.8} (offset: 0x51000) offset reveals cache activity on all letters but no digits.
The library \texttt{libswlo.so} (offset: 0x53e000) shows cache activity on keystrokes reliably, with an F-Score of \SIx{1}. 
Again, inter-keystroke timing attacks can be performed.

\paragrabf{Chrome on Windows.}
On Windows, we tested Chrome versions 103.0.5060.53 and 103.0.5060.114.
Our evaluation runs on a notebook equipped with an Intel i5-4300U running Windows 10 (1803, Build 17134.1726).
We use the \texttt{LoadLibrary} function to load libraries and create a shared mapping between the attacker and victim applications.
Using the key event strings, we directly find the offsets in the DLL file.
We observe that in the \texttt{chrome.dll} (offset: 0xa4ee000) the different key bytes are co-located instead of having a spatial distance of multiple \SI{4}{\kilo\byte} pages.
The Chrome build on Windows uses a different compiler and linker, and, thus, the string merging optimization might behave differently or is even turned off.
Running our cache monitor using \FlushReload we do not observe any cache leakage on the DLL file.
The grouped keys there are KeyA-KeyF,KeyG-KeyS,KeyT-Digit4 and Digit5-Digit9.
Again, the prefetchers on the evaluated CPU practically thwart the distinction between the different groups of keys.
Running a similar experiment as on Linux, we observe a F-Score of \SI{0.99}{\percent}.

In addition, we use our profiler to observe cache activity on other locations.
The DLLs \texttt{msctf.dll} (offset:0x45000)) and \texttt{imm32.dll} (offset:0x3000) also correlate to user input in Chrome on Windows.
These two libraries are mainly used for user input methods and the text management.
Therefore, using \FlushReload inter-keystroke timing attacks are possible on Chrome on Windows.

\paragrabf{Search bar.}
Templating user queries in the browser would tremendously reduce the privacy of browsers.
Running the templater on the search bar of Google Chrome 103.0.5060.53 revealed that the search bar uses a different method to load the keys and there is only a single page (offset: 0x91d4000) in Google Chrome with cache activity upon keystrokes.
Based on our results, we conclude that the search bar does not use the same internal structures for key events as HTML input data.
Still, the leakage we discovered enables inter-keystroke timing attacks on keystrokes.
Running the profiling experiment with all alphanumeric, we achieve a F-Score of \SI{0.99}{\percent} for detecting key presses.

\begin{table*}[t]
  \caption{Evaluated applications. Page cache (PC) and cache-line (CL) indicate whether precise keystroke attacks are possible on that granularity. Inter-Keystroke Timing indicates that key events can be detected on the application via \FlushReload or the page cache.}
  \vspace{-0.1cm}
  \setlength{\aboverulesep}{0pt}
  \setlength{\belowrulesep}{0pt}
  \begin{center}
    \adjustbox{max width=\hsize}{
      \begin{tabular}{llllll}
        \toprule
        Name              & Category               & CL (all key events)    & PC (all key events)    & Inter-Keystroke Timing on Blocks & Avg. F-Score (\FlushReload) \\
        \midrule
        Chrome (99.0.4844.84)         & Browser                & \cmark & \cmark & \cmark & \SI{0.94}{\%}            \\
        Signal-Desktop (5.46.0)   & Private Messenger      & \cmark & \cmark & \cmark & \SI{0.98}{\%}            \\
        Threema (2.4.1)          & Private Messenger      & \cmark & \cmark & \cmark & \SI{0.84}{\%}            \\
        Passky (7.0.0)           & Password Manager       & \cmark & \cmark & \cmark & \SI{0.99}{\%}            \\
        VS-Code (1.69.1)          & Editor                 & \cmark & \cmark & \cmark & \SI{0.85}{\%}            \\
        Chromium Browser (103.0.5060.114) & Browser                & \cmark & \cmark & \cmark & \SI{0.99}{\%}            \\
        Mattermost-Desktop (5.1.1)       & Collaboration Platform & \cmark & \cmark & \cmark & \SI{0.94}{\%}            \\
        Discord (0.0.18)          & Text and Voice Chat    & \cmark & \cmark & \cmark & \SI{0.98}{\%}            \\
        \hline
        Spotify (1.1.84.716) & Audio Streaming        & \cmark & \cmark & \cmark & \SI{0.96}{\%}            \\
        Brackets (1.2.1)  & Editor                    & \xmark & \xmark & \cmark & \SI{0.94}{\%} \\
        \hline
        Chrome 103.0.5060.134(Windows)    & Browser              & \xmark & \xmark & \cmark & \SI{0.99}{\%}            \\
        Chrome 103.0.5060.53 (Search Bar) & Browser              & \xmark & \xmark & \cmark & \SI{0.99}{\%}            \\
        libxul.so (Firefox 102)  & Browser         & \xmark & \xmark & \cmark & \SI{0.99}{\%}            \\
        LibreOffice Writer (6.4.2)      & Office Software        & \xmark & \xmark & \cmark & \SI{0.99}{\%}            \\
        \bottomrule
      \end{tabular}
    }
  \end{center}
  \label{tab:evaluated_programs}
\end{table*}

\section{Mitigation}\label{sec:mitigation}
In this section, we discuss different mitigation vectors that could prevent the leakage.
Nevertheless, we want to emphasize that preventing leakage on such a broad scale as user input in various contexts comes at a significant performance and usability cost.
Furthermore, system-hardening techniques such as ASLR have no effect on the attack as the attacker just maps (which is an unprivileged operation) the victim binary into its own address space.
In principle we identified four conditions for an attack to succeed:
\begin{enumerate}
  \item \textbf{Disable Compiler and Linker Optimizations}:
        For the concrete leakage of Chromium-based applications, disabling the linker optimizations would mitigate accurate keylogging.
        However, it is still possible to detect the activity of key groups as the keys are co-located in \SIx{4} different cache lines.
        Inter-keystroke timing attacks would also enable accurate keylogging as the cache activity can also be used for timing measurements between cache hits~\cite{Song2001,Ristenpart2009,Lipp2016,Naghibijouybari2018,Schwarz2018KeyDrown,Wang2019Unveiling,Paccagnella2021lotr}.
        Moreover, removing these optimizations would bring back again overhead in the linking time ~\cite{Ueyama2019lld}.
  \item \textbf{Golden device availability}:
        All templating attacks require that the attacker is able to run the templating phase on a setup that is similar enough to the victim system~\cite{Chari2002template,Rechberger2004Practical,Brumley2009,Gruss2015Template}.
        This is in principle not difficult to achieve as most desktop and laptop processors today are similar enough with respect to the cache side channel (\ie 64-byte cache lines and the availability of a flush instruction), that it is sufficient to have essentially any desktop or laptop processor available for the templating.
        It is slightly more challenging to obtain precisely the same software version and binary that is running on the target system.
        While software diversity~\cite{Crane2015thwarting} could be an avenue to break this link, it has not really found its way into practice yet.
        In practice, the vast majority of users runs binaries obtained from the official repositories or websites.

        While the idea of templating is decades old already~\cite{Chari2002template}, few works have looked at templating in the context of non-cryptographic leakage from software binaries, \eg via caches~\cite{Gruss2015Template,Wang2019Unveiling}.
        Since our layered approach, \TechniqueName, makes the templating much faster, software diversity could be a possible mitigation to break this condition.
        Without a binary identical to the one the victim runs, \ie due to software diversity, the attacker could not determine templates that work on the victim's system.
  \item \textbf{Secret-dependent execution}:
        The side-channel leakage, \eg which key was pressed, originates in differences in code and data accesses depending on the specific key value.
        The state-of-the-art approach to secure cryptographic code is the linearization to a so-called constant-time implementation.
        Here, code and data accesses are always identical, regardless of the secret processed.
        For cryptographic algorithms this already comes at a non-negligible performance cost~\cite{Chung2012AHigh}.
        On an abstract level, the idea is to always run all the code and access all the data, \eg in a square-and-always-multiply implementation~\cite{Joye2002Montgomery}.
        However, for non-cryptographic algorithms, always running all the code and accessing all the data is entirely infeasible.
        Domas~\cite{DomasMovfuscator2015} presented a compiler that linearizes the entire control flow to a sequence of \texttt{mov} instructions, effectively eliminating all secret-dependent branches, but with an enormous runtime overhead.
        Schwarzl~\etal\cite{Schwarzl2021Specfuscator} further optimized this approach but still observe runtime overheads of factor 1000 and more, even for comparably simple applications.
        Borrello~\etal\cite{Borrello2021Constantine} focused more on the protection of cryptographic implementations and still observe a runtime overhead of factor $3.17$ to $5.07$ on these relatively small examples.
        Hence, the problem of secret dependency on user input in large applications remains an open problem.
  \item \textbf{Side-channel observability}:
        A practical limitation to the exploitation of secret-dependent execution is the side-channel observability.
        While tools like CacheAudit~\cite{Doychev2013CacheAudit} or CaSym~\cite{Brotzman2019casym}, follow the cryptography-focused notion of constant time to consider an application leakage free, the practice for user input is more nuanced.
        For instance, distinguishing keys may be infeasible for an unprivileged attacker in practice when secret key-value-dependent execution exists but does not cross \eg page or cache-line boundaries.
        In particlular, within a page, the hardware prefetcher will pose a substantial obstacle that introduces spurious cache activity on the target cache lines, foiling exploitation in practice~\cite{Gruss2015Template}.

        The compiler could use this effect by grouping potentially secret-dependent accesses closer together or placing these strings in between frequently used code or data.
        We observed this effect in \texttt{gcc} where short strings are directly encoded into the instruction stream, or placed in physical proximity with other strings that are accessed based on a secret value.
        While specific patches can reduce the leakage in the cases we discovered, compilers could in general take side channels into account and employ side-channel-adverse data placement, \ie minimizing the number of cache lines a data structure is spread across.
  \item \textbf{Noise-resilience}:
        Since user input cannot be arbitrarily triggered and repeated by the attacker millions of times, noise-resilience of the side channel is one condition for an attack on user input to succeed~\cite{Schwarz2018KeyDrown}.
        Hence, inducing noise, which is usually not recognized as a viable strategy for cryptographic operations, can provide strong practical security guarantees for user input which occurs at a low frequency and is not arbitrarily repeatable.
        In particular for modern systems, a low number of additional memory accesses for potentially secret dependent operations would come at a negligible performance cost while practically mitigating these attacks on user input.
        Furthermore, user annotations of potentially secret data may help minimizing the performance costs for this approach.
\end{enumerate}

\section{Discussion}\label{sec:discussion}
\TechniqueName is an effective technique to identify leakage in large binaries.
As such, it cannot only used by attackers but is more interesting as a defensive technique revealing leakage already during the development process.
While constant-time implementations have been analyzed for side channels being introduced by compilers~\cite{Page2006A,Van2017Adaptive,Simon2018What,Brennan2020JIT}, developers should be aware that compiler and linker optimizations can lead to spatial distances in binaries enabling side-channel attacks on both hardware and software caches.
We want to emphasize that string deduplication is not only related to native programs. 
Many languages like JavaScript, Java, PHP and Python perform string deduplication (under the term `string interning') to reduce memory utilization.
For instance, Java also offers the possibility of string deduplication (\texttt{-XX:+UseStringDeduplication}) during garbage collection which could lead to similar side effects.

We demonstrated that keystrokes in form input fields in Google Chrome can be detected using cache attacks on hardware and software caches.
While Google Chrome is a valuable target, the dependency of many frameworks on the Chromium project, such as CEF and Electron, leads to a significantly higher impact as hundreds of browser-based desktop applications~\cite{Electron2022Apps} are susceptible to accurate keylogging with our attack.

The templating phase runs in native code on an attacker-controlled system.
While we also demonstrate the exploitation phase only in native code, we want to emphasize that browser-based attacks based on the same leakage are also possible.
Using more advanced techniques like \PrimeProbe in JavaScript and WebAssembly in the browser~\cite{Vila2019theory,Agarwal2022Spookjs}, an attacker may then infer keystrokes from a browser tab, spying on another tab or an Electron app running on the victim's machine.
The crucial challenge of such an attack would be to map the file offsets in the precomputed template to cache sets.
However, \TechniqueName is a generic technique that can be instantiated with various side channels, \eg \PrimeProbe that does not require any shared memory.
Moreover, with techniques by Lipp~\etal\cite{Lipp2016} and Gruss~\etal\cite{Gruss2019page}, \TechniqueName could be applied to Android as well.
\section{Conclusion}\label{sec:conclusion}
We discovered that data deduplication and dead-stripping during compilation and linking facilitate side-channel leakage in compiled binaries.
We show that this effect can even induce side-channel leakage in previously secure binaries, \ie no secret-dependent accesses crossing a 64-byte boundary on the source level but on the binary level.
The foundation to discover this attack was our extension to cache template attacks, called Layered Binary Templating Attacks, \TechniqueName.
\TechniqueName is a scalable approach to templating that combines spatial information from multiple side channels.
Using \TechniqueName we scan binaries compiled with LLVM/clang, which applies deduplication and dead-stripping by default.
Our end-to-end attack is an unprivileged cache-based keylogger for all Chrome-based / Electron-based applications, including hundreds of security-critical apps, \eg the popular Signal messenger app.
While mitigation strategies exist, they come at a cost, and further research is necessary to overcome the open problem of side-channel attacks on user input.
\section*{Acknowledgments}
This work was supported by generous funding and gifts from Red Hat.
We want to thank Hanna Müller, Claudio Canella, Michael Schwarz and Moritz Lipp for valueable feedback on an early draft of this work.
\bibliographystyle{plainurl}
\bibliography{main}

\appendix
\crefalias{section}{appendix}

\section{Cache-hit ratios (extended)}
The cache hit ratio for all lowercase characters with \FlushReload can be seen with~\cref{fig:appendix:ratio_fr} and all alphanumeric characters for the page cache attack~\cref{fig:appendix:ratio_pc}.

\begin{figure*}[t]
  \centering
  \maxsizebox{\hsize}{!}{
    \input{images/ratio_pc.tikz}
  }
  \caption{Cache-hit ratio using a page cache attack for alphanumeric characters in Google Chrome.}
  \label{fig:appendix:ratio_pc}
\end{figure*}
\begin{figure*}[t]
  \centering
  \maxsizebox{\hsize}{!}{
    \input{images/ratio_fr.tikz}
  }
  \caption{Cache-hit ratio using \FlushReload for lowercase letters in Google Chrome.}
  \label{fig:appendix:ratio_fr}
\end{figure*}

\end{document}

%% file: images/attack_overview.tikz
\begin{tikzpicture}

\draw[fill=black!10] (0,-1.5) rectangle +(4.5,) node [pos=.5] {\parbox{2.5cm}{\centering User}};
\draw[] (0,-1.5) rectangle +(4.5,-1.25) node [pos=.5] {\parbox{3.5cm}{}};

\draw[fill=blue!10] (0,-5) rectangle +(4.5,1) node[pos=.5] {\parbox{2.5cm}{\centering Victim}};
\draw[] (0,-5) rectangle +(4.5,-1.25) node [pos=.5] {\parbox{3.5cm}{}};

\draw[fill=red!10] (-5,-1.5) rectangle +(4,1) node[pos=.5] {\parbox{2.5cm}{\centering Attacker}};
\draw(-5,-1.5) rectangle +(4,-1.25) node[pos=.5] {\parbox{2.5cm}{\centering \includegraphics[height=1.25cm]{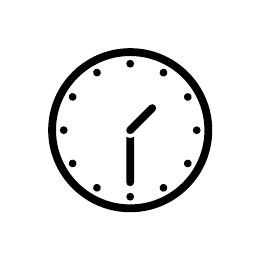}}};

\draw[->,>=stealth,thick] (-3,-2.75) -- (-3,-5) -- node[midway,above] {\textcircled{\raisebox{-1.1pt} {1}} Profile} (0,-5);
\draw[->,>=stealth,thick] (2,-2.75) -- node[midway,xshift=2em,yshift=-0.5em,above] {\textcircled{\raisebox{-1.1pt} {2}} Type}(2,-3.95);
\draw[->,>=stealth,thick] (2,-6.25) -- node[midway,xshift=3.75em,yshift=-0.5em,above] {\textcircled{\raisebox{-1.1pt} {3}} Access cache}(2,-7.5);
\draw[->,>=stealth,thick] (-3.7,-2.75) -- (-3.7,-5) -- (-3.7,-8) to node[midway,above]{\textcircled{\raisebox{-1.1pt} {2}} Monitor} (0,-8);

\draw[fill=orange!10] (0,-8.5) rectangle +(4.5,1) node [pos=.5] {\parbox{2.5cm}{\centering SW/HW Cache}};

{
\node[right] at (-0.1,-2) {\tt \small <input type="password"/>};
\node[right] at (0.35,-5.5) {\tt \small  handleKeystroke();};
}

\end{tikzpicture}

%% file: images/dead_stripping.tikz
\begin{tikzpicture}

\node[right] at (-3.5,-1) {Data};
\node[right] at (-.35,-1) {Compiler};
\node[right] at (3.25,-1) {Binary};

\draw[pattern=north west lines,pattern color=black] (-4,-1.5) rectangle +(2,-5) node [pos=.5] {\parbox{3.5cm}{}};

\draw[] (-0.5,-1.5) rectangle +(2,-5) node [pos=.5] {\parbox{3.5cm}{}};

\draw[fill=green!10] (-4,-1.5) rectangle +(2,-.5) node [pos=.5,xshift=3.5em] {\parbox{3.5cm}{DataA}};
\draw[fill=gray!10] (-4,-2) rectangle +(2,-.5) node [pos=.5,xshift=3.5em] {\parbox{3.5cm}{DataB}};
\draw[fill=gray!10] (-4,-2.5) rectangle +(2,-.5) node [pos=.5,xshift=3.5em] {\parbox{3.5cm}{DataC}};
\draw[fill=blue!10] (-4,-3.) rectangle +(2,-.5) node [pos=.5,xshift=3.5em] {\parbox{3.5cm}{DataD}};
\draw[fill=gray!10] (-4,-3.5) rectangle +(2,-.5) node [pos=.5,xshift=3.5em] {\parbox{3.5cm}{DataE}};
\draw[fill=gray!10] (-4,-4) rectangle +(2,-.5) node [pos=.5,xshift=3.5em] {\parbox{3.5cm}{DataF}};
\draw[fill=gray!10] (-4,-4.5) rectangle +(2,-.5) node [pos=.5,xshift=3.5em] {\parbox{3.5cm}{DataG}};
\draw[fill=red!10] (-4,-5) rectangle +(2,-.5) node [pos=.5,xshift=3.5em] {\parbox{3.5cm}{DataH}};
\draw[fill=gray!10] (-4,-5.5) rectangle +(2,-.5) node [pos=.5,xshift=3.5em] {\parbox{3.5cm}{DataI}};
\draw[fill=gray!10] (-4,-6) rectangle +(2,-.5) node [pos=.5,xshift=3.5em] {\parbox{3.5cm}{DataJ}};

\draw[pattern=north west lines,pattern color=black] (3,-1.5) rectangle +(2,-1.25) node [pos=.5] {\parbox{3.5cm}{}};
\draw[pattern=north west lines,pattern color=black] (3,-2.75) rectangle +(2,-1.25) node [pos=.5] {\parbox{3.5cm}{}};
\draw[pattern=north west lines,pattern color=black] (3,-4) rectangle +(2,-1.25) node [pos=.5] {\parbox{3.5cm}{}};
\draw[pattern=north west lines,pattern color=black] (3,-5.25) rectangle +(2,-1.25) node [pos=.5] {\parbox{3.5cm}{}};

\draw[fill=green!10] (3,-1.75) rectangle +(2,-.5) node [pos=.5,xshift=3.5em] {\parbox{3.5cm}{DataA}};
\draw[fill=blue!10] (3,-2.25) rectangle +(2,-.5) node [pos=.5,xshift=3.5em] {\parbox{3.5cm}{DataD}};
\draw[fill=red!10] (3,-2.75) rectangle +(2,-.5) node [pos=.5,xshift=3.5em] {\parbox{3.5cm}{DataH}};

\node[right,rotate=-90] at (.5,-3) {Place data};

\draw[<->] (-2,-1.75) to node[midway,yshift=0.4em] {used} (-0.5,-1.75);
\draw[<->] (-2,-2.25) to node[midway,yshift=0.4em] {unused} (-0.5,-2.25);
\draw[<->] (-2,-2.75) to node[midway,yshift=0.4em] {unused}(-0.5,-2.75);
\draw[<->] (-2,-3.25) to node[midway,yshift=0.4em] {used}(-0.5,-3.25);
\draw[<->] (-2,-3.75) to node[midway,yshift=0.4em] {unused}(-0.5,-3.75);
\draw[<->] (-2,-4.25) to node[midway,yshift=0.4em] {unused}(-0.5,-4.25);
\draw[<->] (-2,-4.75) to node[midway,yshift=0.4em] {unused}(-0.5,-4.75);
\draw[<->] (-2,-5.25) to node[midway,yshift=0.4em] {used}(-0.5,-5.25);
\draw[<->] (-2,-5.75) to node[midway,yshift=0.4em] {unused}(-0.5,-5.75);
\draw[<->] (-2,-6.25) to node[midway,yshift=0.4em] {unused}(-0.5,-6.25);

\draw[->] (1.5,-2) to (2.95,-2);
\draw[->] (1.5,-2.5) to (2.95,-2.5);
\draw[->] (1.5,-3) to (2.95,-3);

\end{tikzpicture}

%% file: images/compiler_string_merging.tikz
\begin{tikzpicture}

\node[right] at (-3.5,-1) {Data};
\node[right] at (-.35,-1) {Compiler};
\node[right] at (3.25,-1) {Binary};

\draw[pattern=north west lines,pattern color=black] (-4,-1.5) rectangle +(2,-5) node [pos=.5] {\parbox{3.5cm}{}};

\draw[] (-0.5,-1.5) rectangle +(2,-5) node [pos=.5] {\parbox{3.5cm}{}};

\draw[fill=green!10] (-4,-1.5) rectangle +(2,-.5) node [pos=.5,xshift=3.5em] {\parbox{3.5cm}{KeyA}};
\draw[fill=blue!10] (-4,-2) rectangle +(2,-.5) node [pos=.5,xshift=3.5em] {\parbox{3.5cm}{KeyB}};
\draw[fill=red!10] (-4,-2.5) rectangle +(2,-.5) node [pos=.5,xshift=3.5em] {\parbox{3.5cm}{KeyC}};
\draw[fill=orange!10] (-4,-3.) rectangle +(2,-.5) node [pos=.5,xshift=3.5em] {\parbox{3.5cm}{KeyD}};

\draw[fill=green!10] (-4,-4) rectangle +(2,-.5) node [pos=.5,xshift=3.5em] {\parbox{3.5cm}{KeyA}};
\draw[fill=blue!10] (-4,-4.5) rectangle +(2,-.5) node [pos=.5,xshift=3.5em] {\parbox{3.5cm}{KeyB}};
\draw[fill=red!10] (-4,-5) rectangle +(2,-.5) node [pos=.5,xshift=3.5em] {\parbox{3.5cm}{KeyC}};
\draw[fill=orange!10] (-4,-5.5) rectangle +(2,-.5) node [pos=.5,xshift=3.5em] {\parbox{3.5cm}{KeyD}};

\draw[pattern=north west lines,pattern color=black] (3,-1.5) rectangle +(2,-1.25) node [pos=.5] {\parbox{3.5cm}{}};
\draw[pattern=north west lines,pattern color=black] (3,-2.75) rectangle +(2,-1.25) node [pos=.5] {\parbox{3.5cm}{}};
\draw[pattern=north west lines,pattern color=black] (3,-4) rectangle +(2,-1.25) node [pos=.5] {\parbox{3.5cm}{}};
\draw[pattern=north west lines,pattern color=black] (3,-5.25) rectangle +(2,-1.25) node [pos=.5] {\parbox{3.5cm}{}};

\draw[fill=green!10] (3,-1.75) rectangle +(2,-.5) node [pos=.5,xshift=3.5em] {\parbox{3.5cm}{KeyA}};
\draw[fill=blue!10] (3,-3.25) rectangle +(2,-.5) node [pos=.5,xshift=3.5em] {\parbox{3.5cm}{KeyB}};
\draw[fill=red!10] (3,-4.5) rectangle +(2,-.5) node [pos=.5,xshift=3.5em] {\parbox{3.5cm}{KeyC}};
\draw[fill=orange!10] (3,-5.5) rectangle +(2,-.5) node [pos=.5,xshift=3.5em] {\parbox{3.5cm}{KeyD}};

\node[right] at (5,-2.75) {\scriptsize 0x21000};
\node[right] at (5,-4) {\scriptsize 0x22000};
\node[right] at (5,-5.25) {\scriptsize 0x23000};
\node[right] at (5,-6.5) {\scriptsize 0x24000};

\node[right,rotate=-90] at (.5,-2.25) {Deduplicate strings};

\draw[<->] (-2,-1.75) to (-0.5,-1.75);
\draw[<->,in=175] (-2,-4.25) to (-0.5,-2);

\draw[<->] (-2,-2.25) to (-0.5,-2.75);
\draw[<->,in=175] (-2,-4.75) to (-0.5,-3);

\draw[<->] (-2,-2.75) to (-0.5,-3.75);
\draw[<->,in=175] (-2,-5.25) to (-0.5,-4);

\draw[<->] (-2,-3.25) to (-0.5,-4.75);
\draw[<->,in=175] (-2,-5.75) to (-0.5,-5);

\draw[<->] (1.5,-1.75) to (2.95,-2);
\draw[<->] (1.5,-2.75) to (2.95,-3.5);
\draw[<->] (1.5,-3.75) to (2.95,-4.75);
\draw[<->] (1.5,-4.75) to (2.95,-5.75);

\end{tikzpicture}

%% file: images/linker_string_merging.tikz
\begin{tikzpicture}

\node[right] at (-3.5,-1) {Data};
\node[right] at (-.55,-1) {Hash-Tables};
\node[right] at (3.25,-1) {Binary};

\draw[pattern=north west lines,pattern color=black] (-4,-1.5) rectangle +(2,-5) node [pos=.5] {\parbox{3.5cm}{}};

\draw[fill=green!10] (-4,-1.5) rectangle +(2,-.5) node [pos=.5,xshift=3.5em] {\parbox{3.5cm}{KeyA}};
\draw[fill=blue!10] (-4,-2) rectangle +(2,-.5) node [pos=.5,xshift=3.5em] {\parbox{3.5cm}{KeyB}};
\draw[fill=red!10] (-4,-2.5) rectangle +(2,-.5) node [pos=.5,xshift=3.5em] {\parbox{3.5cm}{KeyC}};
\draw[fill=orange!10] (-4,-3.) rectangle +(2,-.5) node [pos=.5,xshift=3.5em] {\parbox{3.5cm}{KeyD}};

\draw[fill=green!10] (-4,-4) rectangle +(2,-.5) node [pos=.5,xshift=3.5em] {\parbox{3.5cm}{KeyA}};
\draw[fill=blue!10] (-4,-4.5) rectangle +(2,-.5) node [pos=.5,xshift=3.5em] {\parbox{3.5cm}{KeyB}};
\draw[fill=red!10] (-4,-5) rectangle +(2,-.5) node [pos=.5,xshift=3.5em] {\parbox{3.5cm}{KeyC}};
\draw[fill=orange!10] (-4,-5.5) rectangle +(2,-.5) node [pos=.5,xshift=3.5em] {\parbox{3.5cm}{KeyD}};

\draw[fill=green!10] (-0.5,-2) rectangle +(2,0.5) node(HT1) [pos=.5,xshift=3.5em] {\parbox{3.5cm}{HT 1}};
\draw[fill=blue!10] (-0.5,-3) rectangle +(2,0.5) node [pos=.5,xshift=3.5em] {\parbox{3.5cm}{HT 2}};
\draw[fill=red!10] (-0.5,-4) rectangle +(2,0.5) node [pos=.5,xshift=3.5em] {\parbox{3.5cm}{HT 3}};
\draw[fill=orange!10] (-0.5,-5) rectangle +(2,0.5) node [pos=.5,xshift=3.5em] {\parbox{3.5cm}{HT 3}};

\draw[pattern=north west lines,pattern color=black] (3,-1.5) rectangle +(2,-1.25) node [pos=.5] {\parbox{3.5cm}{}};
\draw[pattern=north west lines,pattern color=black] (3,-2.75) rectangle +(2,-1.25) node [pos=.5] {\parbox{3.5cm}{}};
\draw[pattern=north west lines,pattern color=black] (3,-4) rectangle +(2,-1.25) node [pos=.5] {\parbox{3.5cm}{}};
\draw[pattern=north west lines,pattern color=black] (3,-5.25) rectangle +(2,-1.25) node [pos=.5] {\parbox{3.5cm}{}};

\draw[fill=green!10] (3,-1.75) rectangle +(2,-.5) node [pos=.5,xshift=3.5em] {\parbox{3.5cm}{KeyA}};
\draw[fill=blue!10] (3,-3.25) rectangle +(2,-.5) node [pos=.5,xshift=3.5em] {\parbox{3.5cm}{KeyB}};
\draw[fill=red!10] (3,-4.5) rectangle +(2,-.5) node [pos=.5,xshift=3.5em] {\parbox{3.5cm}{KeyC}};
\draw[fill=orange!10] (3,-5.5) rectangle +(2,-.5) node [pos=.5,xshift=3.5em] {\parbox{3.5cm}{KeyD}};

\node[right] at (5,-2.75) {\scriptsize 0x21000};
\node[right] at (5,-4) {\scriptsize 0x22000};
\node[right] at (5,-5.25) {\scriptsize 0x23000};
\node[right] at (5,-6.5) {\scriptsize 0x24000};

\draw[<->] (-2,-1.75) to (-0.5,-1.75);
\draw[<->,in=175] (-2,-4.25) to (-0.5,-2);

\draw[<->] (-2,-2.25) to (-0.5,-2.75);
\draw[<->,in=175] (-2,-4.75) to (-0.5,-3);

\draw[<->] (-2,-2.75) to (-0.5,-3.75);
\draw[<->,in=175] (-2,-5.25) to (-0.5,-4);

\draw[<->] (-2,-3.25) to (-0.5,-4.75);
\draw[<->,in=175] (-2,-5.75) to (-0.5,-5);

\draw[->] (1.5,-1.75) to node [pos=.5,midway,yshift=+0.75em] {Write} (2.95,-2);
\draw[->] (1.5,-2.75) to (2.95,-3.5);
\draw[->] (1.5,-3.75) to (2.95,-4.75);
\draw[->] (1.5,-4.75) to (2.95,-5.75);

\end{tikzpicture}

%% file: images/chrome_leakage_rodata.tikz
\begin{tikzpicture}

\draw[fill=black!10] (0,-1.5) rectangle +(2,1) node [pos=.5] {\parbox{2.5cm}{\centering .rodata}};
\draw[] (0,-1.5) rectangle +(2,-1.25) node [pos=.5] {\parbox{3.5cm}{}};
\draw[] (0,-2.75) rectangle +(2,-1.25) node [pos=.5] {\parbox{3.5cm}{}};
\draw[] (0,-4) rectangle +(2,-1.25) node [pos=.5] {\parbox{3.5cm}{}};
\draw[] (0,-5.25) rectangle +(2,-1.25) node [pos=.5] {\parbox{3.5cm}{}};

\draw[fill=green!10] (0,-1.75) rectangle +(2,-.5) node [pos=.5,xshift=3.5em] {\parbox{3.5cm}{KeyA}};
\draw[fill=blue!10] (0,-3.25) rectangle +(2,-.5) node [pos=.5,xshift=3.5em] {\parbox{3.5cm}{KeyB}};
\draw[fill=red!80] (0,-4.5) rectangle +(2,-.5) node [pos=.5,xshift=3.5em] {\parbox{3.5cm}{KeyC}};
\draw[fill=orange!10] (0,-5.5) rectangle +(2,-.5) node [pos=.5,xshift=3.5em] {\parbox{3.5cm}{KeyD}};

\draw[fill=red!10] (-5,-1.5) rectangle +(4,1) node[pos=.5] {\parbox{2.5cm}{\centering User}};
\draw(-5,-1.5) rectangle +(4,-1.25) node[pos=.5] {};

\draw[->,>=stealth,thick] (-3,-2.75) -- (-3,-4.75) -- node[midway,above] {Type C} (0,-4.75);
\draw[>=stealth,thick] (2,-4.75) -- node[midway,above] {\scriptsize Page / Cache Line accessed} (6,-4.75);

\end{tikzpicture}

%% file: images/ratio_digits_fr.tikz
\begin{tikzpicture}[]
  \foreach \y [count=\n] in {
      {115,0,0,0,0,0,1,0,0,43},
      {5,185,0,0,0,0,0,0,0,47},
      {0,0,165,0,0,0,0,0,0,53},
      {2,0,0,178,0,0,0,0,0,45},
      {0,0,0,0,171,0,0,0,0,49},
      {0,0,0,0,0,173,0,0,0,56},
      {0,0,0,0,0,0,156,0,0,51},
      {0,0,0,0,0,0,0,169,0,58},
      {0,0,0,0,0,0,0,0,165,52},
      {0,0,0,0,0,0,0,0,0,140},
    } {
      \foreach \n in {0,1,2,3,4,5,6,7,8,9}
      {
        \node[minimum size=6mm] at (\n + 1, 0) {\n};
      }
      \foreach \x [count=\m] in \y {
        \ifthenelse{\x < 60}{%
          \node[text=black,fill=black!\x!white, minimum size=6mm] at (\m,-\n) {\x};%
        }{%
          \node[text=white,fill=black!\x!white, minimum size=6mm] at (\m,-\n) {\x};%
        }
      }
    }

  \foreach \a [count=\i] in {0x1521203,0x151b9bd,0x1513d05,0x1510118,0x150d59a,0x150b526,0x1509a35,0x1508991,0x1506eb8,0x1505e5b} {
    \node[minimum size=10mm,left] at (0,-\i) {\a};
  }
\end{tikzpicture}

%% file: images/fscore_fr.tikz
\begin{tikzpicture}
\begin{axis}[
style={font=\footnotesize},
xlabel={Key Event},
ylabel={F1-Score},
y label style={align=center,text width=2cm},
width=\hsize,
height=4.25cm,
xtick={0,1,2,3,4,5,6,7,8,9,10,11,12,13,14,15,16,17,18,19,20,21,22,23,24,25,26,27,28,29,30,31,32,33,34,35},
xticklabels={a,b,c,d,e,f,g,h,i,j,k,l,m,n,o,p,q,r,s,t,u,v,w,x,y,z,0,1,2,3,4,5,6,7,8,9},
xticklabel style={text height=1ex},
]

\addplot+[mark=*,draw=none] table[x index=1,y index=2,col sep=comma] {data/f_score_fr.csv};
\end{axis}
\end{tikzpicture}

%% file: images/fscore_pc.tikz
\begin{tikzpicture}
\begin{axis}[
style={font=\footnotesize},
xlabel={Key Event},
ylabel={F1-Score},
y label style={align=center,text width=2cm},
width=\hsize,
height=4.25cm,
xtick={0,1,2,3,4,5,6,7,8,9,10,11,12,13,14,15,16,17,18,19,20,21,22,23,24,25,26,27,28,29,30,31,32,33,34,35},
xticklabels={a,b,c,d,e,f,g,h,i,j,k,l,m,n,o,p,q,r,s,t,u,v,w,x,y,z,0,1,2,3,4,5,6,7,8,9},
xticklabel style={text height=1ex},
]

\addplot+[mark=*,draw=none] table[x index=1,y index=2,col sep=comma] {data/f_score_pc.csv};
\end{axis}
\end{tikzpicture}

%% file: images/ratio_pc.tikz
\resizebox{0.8\hsize}{!}{
\begin{tikzpicture}[]
  \foreach \y [count=\n] in {
      {a,c,d,e,f,g,i,j,k,l,n,o,p,q,r,s,t,u,v,w,z,0,1,2,3,4,5,6,7,8,bm9hyx},
    {98,0,0,0,0,0,0,0,0,0,0,0,0,0,0,0,0,0,0,0,0,0,0,0,0,0,0,0,0,0,2},
    {0,0,0,0,0,0,0,0,0,0,0,0,0,0,0,0,0,0,0,0,0,0,0,0,0,0,0,0,0,0,100},
    {0,95,0,0,0,0,0,0,0,0,0,0,0,0,0,0,0,0,0,0,0,0,0,0,0,0,0,0,0,0,0},
    {0,0,99,0,0,0,0,0,0,0,0,0,0,0,0,0,0,0,0,0,0,0,0,0,0,0,0,0,0,0,0},
    {0,0,0,98,0,0,0,0,0,0,0,0,0,0,0,0,0,0,0,0,0,0,0,0,0,0,0,0,0,0,1},
    {0,0,0,0,97,0,0,0,0,0,0,0,0,0,0,0,0,0,0,0,0,0,0,0,0,0,0,0,0,0,0},
    {0,0,0,0,0,98,0,1,0,0,0,0,0,0,0,0,0,0,0,0,0,0,0,0,0,0,0,0,0,0,1},
    {0,0,0,0,0,0,0,0,0,0,0,0,0,0,0,0,0,0,0,0,0,0,0,0,0,0,0,0,0,0,100},
    {0,0,0,0,0,0,99,0,0,0,0,0,0,0,0,0,0,0,0,0,0,0,0,0,0,0,0,0,0,0,0},
    {0,0,0,0,0,0,0,100,0,0,0,0,0,0,0,0,0,0,0,0,0,0,0,0,0,0,0,0,0,0,0},
    {0,0,0,0,0,0,0,0,98,0,0,0,0,0,0,0,0,0,0,0,0,0,0,0,0,0,0,0,0,0,2},
    {0,0,0,0,0,0,0,0,0,91,0,0,0,0,0,0,0,0,0,0,0,0,0,0,0,0,0,0,0,0,6},
    {0,0,0,0,0,0,0,0,0,0,0,0,0,0,0,0,0,0,0,0,0,0,0,0,0,0,0,0,0,0,100},
    {0,0,0,0,0,0,0,0,0,0,97,0,0,0,0,0,0,0,0,0,0,0,0,0,0,0,0,0,0,0,2},
    {0,0,0,0,0,0,0,0,0,0,0,94,0,0,0,0,0,0,0,0,0,0,0,0,0,0,0,0,0,0,1},
    {0,0,0,0,0,0,0,0,0,0,0,0,99,0,0,0,0,0,0,0,0,0,0,0,0,0,0,0,0,0,1},
    {0,0,0,0,0,0,0,0,0,0,0,0,0,100,0,0,0,0,0,0,0,0,0,0,0,0,0,0,0,0,0},
    {0,0,0,0,0,0,0,0,0,0,0,0,0,0,97,0,0,0,0,0,0,0,0,0,0,0,0,0,0,0,1},
    {0,0,0,0,0,0,0,0,0,0,0,0,0,0,0,97,0,0,0,0,0,0,0,0,0,0,0,0,0,0,1},
    {0,0,0,0,0,0,0,0,0,0,0,0,0,0,0,0,96,0,0,0,0,0,0,0,0,0,0,0,0,0,2},
    {0,0,0,0,0,0,0,0,0,0,0,0,0,0,0,0,0,98,0,0,0,0,0,0,0,0,0,0,0,0,1},
    {0,0,0,0,0,0,0,0,0,0,0,0,0,0,0,0,0,0,93,0,0,0,0,0,0,0,0,0,0,0,1},
    {0,0,0,0,0,0,0,0,0,0,0,0,0,0,0,0,0,0,0,99,0,0,0,0,0,0,0,0,0,0,1},
    {0,0,0,0,0,0,0,0,0,0,0,0,0,0,0,0,0,0,0,0,0,0,0,0,0,0,0,0,0,0,100},
    {0,0,0,0,0,0,0,0,0,0,0,0,0,0,0,0,0,0,0,0,0,0,0,0,0,0,0,0,0,0,100},
    {0,0,0,0,0,0,0,0,0,0,0,0,0,0,0,0,0,0,0,0,98,0,0,0,0,0,0,0,0,0,1},
    {0,0,0,0,0,0,0,0,0,0,0,0,0,0,0,0,0,0,0,0,0,98,0,0,0,0,0,0,0,0,1},
    {0,0,0,0,0,0,0,0,0,0,0,0,0,0,0,0,0,0,0,0,0,0,99,0,0,0,0,0,0,0,1},
    {0,0,0,0,0,0,0,0,0,0,0,0,0,0,0,0,0,0,0,0,0,0,0,97,0,0,0,0,0,0,3},
    {0,0,0,0,0,0,0,0,0,0,0,0,0,0,0,0,0,0,0,0,0,0,0,0,98,0,0,0,0,0,1},
    {0,0,0,0,0,0,0,0,0,0,0,0,0,0,0,0,0,0,0,0,0,0,0,0,0,97,0,0,0,0,1},
    {0,0,0,0,0,0,0,0,0,0,0,0,0,0,0,0,0,0,0,0,0,0,0,0,0,0,97,0,0,0,2},
    {0,0,0,0,0,0,0,0,0,0,0,0,0,0,0,0,0,0,0,0,0,0,0,0,0,0,0,98,0,0,0},
    {0,0,0,0,0,0,0,0,0,0,0,0,0,0,0,0,0,0,0,0,0,0,0,0,0,0,0,0,97,0,0},
    {0,0,0,0,0,0,0,0,0,0,0,0,0,0,0,0,0,0,0,0,0,0,0,0,0,0,0,0,0,100,0},
    {0,0,0,0,0,0,0,0,0,0,0,0,0,0,0,0,0,0,0,0,0,0,0,0,0,0,0,0,0,0,100},
    } {
      \ifnum\n<2
        \foreach \x [count=\m] in \y {
          \node[minimum size=6mm] at (\m, 0) {\x};
        }
      \fi
      \ifnum\n>1
      \foreach \x [count=\m] in \y {
        \ifthenelse{\x < 60}{%
          \node[text=black,fill=black!\x!white, minimum size=6mm] at (\m,-\n+1) {\x};%
        }{%
          \node[text=white,fill=black!\x!white, minimum size=6mm] at (\m,-\n+1) {\x};%
        }
      }
      \fi
    }

  \foreach \a [count=\i] in {0x14e7000,0x14e2000,0x14df000,0x14d2000,0x14c1000,0x14c0000,0x14be000,0x14bc000,0x14bb000,0x14ba000,0x14b9000,0x14b3000,0x14af000,0x14ab000,0x14aa000,0x14a9000,0x14a8000,0x14a2000,0x149b000,0x1494000,0x1492000,0x148f000,0x148e000,0x148c000,0x148b000,0x148a000,0x1521000,0x151b000,0x1513000,0x1510000,0x150d000,0x150b000,0x1509000,0x1508000,0x1506000,0x1505000} {
    \node[minimum size=10mm,left] at (0,-\i) {\a};
  }
\end{tikzpicture}
}

%% file: images/ratio_fr.tikz
\begin{tikzpicture}[]
  \foreach \y [count=\n] in {
      {a,b,c,d,e,f,g,h,i,j,k,l,m,n,o,p,q,r,s,t,u,v,w,x,y,z},
      {99,1,0,0,0,0,0,1,0,0,0,0,0,0,0,0,2,0,0,2,0,0,8,7,0,0},
      {2,178,0,0,0,0,0,0,0,0,0,0,0,0,0,0,0,0,0,0,0,0,0,0,0,0},
      {0,0,151,0,0,0,0,0,0,0,0,0,0,0,0,0,0,0,0,0,0,0,0,0,0,0},
      {0,0,0,173,0,0,0,0,0,0,0,0,0,0,0,0,0,0,0,0,0,0,0,0,0,0},
      {0,1,0,0,101,0,0,0,0,0,0,0,0,0,0,0,0,0,0,0,0,0,0,0,0,0},
      {0,0,0,0,0,100,0,0,0,0,0,0,0,0,0,0,0,0,0,0,0,0,0,0,0,0},
      {0,0,0,0,0,0,96,0,0,0,0,0,0,0,0,0,0,0,0,0,0,1,0,0,0,0},
      {0,0,0,0,0,0,0,169,0,0,0,0,0,0,0,0,0,0,0,0,0,0,0,0,0,0},
      {0,0,0,0,0,0,0,0,114,0,0,0,0,0,0,0,0,0,0,0,0,0,0,0,0,0},
      {0,0,0,0,0,0,0,0,0,156,0,0,0,0,0,0,0,0,0,0,0,0,0,0,0,0},
      {0,1,0,0,0,0,0,0,0,0,171,0,0,0,0,0,0,0,0,0,0,0,0,0,0,0},
      {0,0,0,0,0,0,0,0,0,0,0,146,0,0,0,0,0,0,0,0,0,0,0,0,0,0},
      {0,1,0,0,0,0,0,0,0,0,0,0,125,0,0,0,0,0,0,0,0,0,0,0,0,0},
      {0,0,0,0,0,0,0,0,0,0,0,0,0,148,0,0,0,0,0,0,0,1,0,0,0,0},
      {0,0,0,0,0,0,0,0,0,0,0,0,0,0,165,0,0,0,0,0,0,0,0,0,0,0},
      {0,0,0,0,0,0,0,0,0,0,0,0,0,0,0,87,0,0,0,0,0,0,0,0,0,0},
      {0,0,0,0,0,0,0,0,0,0,0,0,0,0,0,15,156,0,0,0,0,0,0,0,0,0},
      {0,1,0,0,0,0,0,0,0,0,0,0,0,0,0,0,0,141,0,0,0,0,0,0,0,0},
      {0,0,0,0,0,0,0,0,0,0,0,0,0,0,0,0,0,0,108,0,0,0,0,0,0,0},
      {0,0,0,0,0,0,0,0,0,0,0,0,0,0,0,0,0,0,0,157,0,0,0,0,0,0},
      {0,1,0,0,0,0,0,0,0,0,0,0,0,0,0,0,0,0,0,0,135,0,0,0,0,0},
      {0,0,0,0,0,0,0,0,0,0,0,0,0,0,0,0,0,0,0,0,0,107,0,0,0,0},
      {0,0,0,0,0,0,0,0,0,0,0,0,0,0,0,0,0,0,0,0,0,0,124,0,0,0},
      {0,0,0,0,0,0,0,0,0,0,0,0,0,0,0,0,0,0,0,0,0,0,0,138,0,0},
      {0,0,0,0,0,0,0,0,0,0,0,0,0,0,0,0,0,0,0,0,0,0,0,0,137,0},
      {0,0,0,0,0,0,0,0,0,0,0,0,0,0,0,0,0,0,0,0,0,0,0,0,0,141},
    } {
      \ifnum\n<2
        \foreach \x [count=\m] in \y {
          \node[minimum size=6mm] at (\m, 0) {\x};
        }
      \fi
      \ifnum\n>1
      \foreach \x [count=\m] in \y {
        \ifthenelse{\x < 60}{%
          \node[text=black,fill=black!\x!white, minimum size=6mm] at (\m,-\n+1) {\x};%
        }{%
          \node[text=white,fill=black!\x!white, minimum size=6mm] at (\m,-\n+1) {\x};%
        }
      }
      \fi
    }

  \foreach \a [count=\i] in {0x14e7e5b,0x14e24a8,0x14df043,0x14d2ab9,0x14c1a5e,0x14c0a76,0x14be05f,0x14bc5ad,0x14bb06c,0x14bae3b,0x14b981f,0x14b3350,0x14af573,0x14ab755,0x14aa938,0x14a9172,0x14a8e1c,0x14a2f87,0x149b36d,0x149411c,0x1492c1d,0x148f151,0x148e546,0x148cb27,0x148b08b,0x148ac19} {
    \node[minimum size=10mm,left] at (0,-\i) {\a};
  }
\end{tikzpicture}

%% file: main.bbl
\begin{thebibliography}{100}

\bibitem{Agarwal2022Spookjs}
Ayush Agarwal, Sioli O'Connell, Jason Kim, Shaked Yehezkel, Daniel Genkin, Eyal
  Ronen, and Yuval Yarom.
\newblock {Spook.js: Attacking Chrome Strict Site Isolation via Speculative
  Execution}.
\newblock In {\em S\&P}, 2022.

\bibitem{Anonymous2022blind}
{Anonymous}.
\newblock {Blinded for Peer Review}, 2022.

\bibitem{Korpi2022xkbcat}
{Antti Korpi}.
\newblock {xkbcat}, 2021.
\newblock URL: \url{https://github.com/anko/xkbcat}.

\bibitem{bacs2022dupefs}
Andrei Bacs, Saidgani Musaev, Kaveh Razavi, Cristiano Giuffrida, and Herbert
  Bos.
\newblock {DUPEFS}: {Leaking Data Over the Network With Filesystem
  Deduplication Side Channels}.
\newblock In {\em FAST}, 2022.

\bibitem{Github2022wayland}
Maarten Baert.
\newblock {wayland-keylogger}, 2022.
\newblock URL: \url{https://github.com/Aishou/wayland-keylogger}.

\bibitem{Barresi2015}
Antonio Barresi, Kaveh Razavi, Mathias Payer, and Thomas~R. Gross.
\newblock {CAIN:} silently breaking {ASLR} in the cloud.
\newblock In {\em {WOOT}}, 2015.

\bibitem{Bernstein2005}
Daniel~J. Bernstein.
\newblock {Cache-Timing Attacks on AES}, 2005.
\newblock URL: \url{http://cr.yp.to/antiforgery/cachetiming-20050414.pdf}.

\bibitem{Borrello2021Constantine}
Pietro Borrello, Daniele~Cono D'Elia, Leonardo Querzoni, and Cristiano
  Giuffrida.
\newblock {Constantine: Automatic Side-Channel Resistance Using Efficient
  Control and Data Flow Linearization}.
\newblock In {\em CCS}, 2021.

\bibitem{Bosman2016}
Erik Bosman, Kaveh Razavi, Herbert Bos, and Cristiano Giuffrida.
\newblock {Dedup Est Machina: Memory Deduplication as an Advanced Exploitation
  Vector}.
\newblock In {\em S\&P}, 2016.

\bibitem{Brasser2017sgx}
Ferdinand Brasser, Urs M{\"u}ller, Alexandra Dmitrienko, Kari Kostiainen,
  Srdjan Capkun, and Ahmad-Reza Sadeghi.
\newblock {Software Grand Exposure: SGX Cache Attacks Are Practical}.
\newblock In {\em {WOOT}}, 2017.

\bibitem{Brennan2020JIT}
Tegan Brennan, Nicol{\'a}s Rosner, and Tevfik Bultan.
\newblock {JIT Leaks: inducing timing side channels through just-in-time
  compilation}.
\newblock In {\em S\&P}, 2020.

\bibitem{Brotzman2019casym}
Robert Brotzman, Shen Liu, Danfeng Zhang, Gang Tan, and Mahmut Kandemir.
\newblock {CaSym: Cache aware symbolic execution for side channel detection and
  mitigation}.
\newblock In {\em S\&P}, 2019.

\bibitem{Brumley2009}
Billy Brumley and Risto Hakala.
\newblock {Cache-Timing Template Attacks}.
\newblock In {\em AsiaCrypt}, 2009.

\bibitem{Bruno2013technet}
Luigi Bruno.
\newblock {What page replacement algorithms does the windows 7 OS uses?},
  February 2013.
\newblock URL:
  \url{https://social.technet.microsoft.com/Forums/ie/en-US/e61aef24-38fd-4e7e-a4c1-a50aa226818c}.

\bibitem{Carre2019End-to-end}
Sebastien Carre, Victor Dyseryn, Adrien Facon, Sylvain Guilley, and Thomas
  Perianin.
\newblock {End-to-end automated cache-timing attack driven by Machine
  Learning}.
\newblock {\em Journal of Cryptology}, 2019.

\bibitem{Cauligi2020towards}
Sunjay Cauligi, Craig Disselkoen, Klaus~v Gleissenthall, Deian Stefan, Tamara
  Rezk, and Gilles Barthe.
\newblock {Towards Constant-Time Foundations for the New Spectre Era}.
\newblock In {\em PLDI}, 2020.

\bibitem{Cauligi2017Fact}
Sunjay Cauligi, Gary Soeller, Fraser Brown, Brian Johannesmeyer, Yunlu Huang,
  Ranjit Jhala, and Deian Stefan.
\newblock {FaCT: A flexible, constant-time programming language}.
\newblock In {\em SecDev}, 2017.

\bibitem{CEF2022Chrome}
{CEF}.
\newblock {Chrome Embedded Framework}, 2022.
\newblock URL: \url{https://github.com/chromiumembedded/cef}.

\bibitem{Chari2002template}
Suresh Chari, Josyula~R Rao, and Pankaj Rohatgi.
\newblock {Template attacks}.
\newblock In {\em {CHES}}, 2002.

\bibitem{Chromium2022Speeding}
{Chromium}.
\newblock {Speeding up Chrome's release cycle}, 2022.
\newblock URL:
  \url{https://blog.chromium.org/2021/03/speeding-up-release-cycle.html}.

\bibitem{Chung2012AHigh}
Szu-Chi Chung, Jen-Wei Lee, Hsie-Chia Chang, and Chen-Yi Lee.
\newblock {A high-performance elliptic curve cryptographic processor over GF(p)
  with SPA resistance}.
\newblock In {\em International Symposium on Circuits and Systems (ISCAS)},
  2012.

\bibitem{Coppens2009}
Bart Coppens, Ingrid Verbauwhede, Koen {De Bosschere}, and Bjorn {De Sutter}.
\newblock {Practical mitigations for timing-based side-channel attacks on
  modern x86 processors}.
\newblock In {\em S\&P}, 2009.

\bibitem{Costi2022On}
Andreas Costi, Brian Johannesmeyer, Erik Bosman, Cristiano Giuffrida, and
  Herbert Bos.
\newblock On the effectiveness of same-domain memory deduplication.
\newblock In {\em European Workshop on Systems Security}, pages 29--35, 2022.

\bibitem{Crane2015thwarting}
Stephen Crane, Andrei Homescu, Stefan Brunthaler, Per Larsen, and Michael
  Franz.
\newblock {Thwarting Cache Side-Channel Attacks Through Dynamic Software
  Diversity}.
\newblock In {\em NDSS}, 2015.

\bibitem{Dall2018cachequote}
Fergus Dall, Gabrielle De~Micheli, Thomas Eisenbarth, Daniel Genkin, Nadia
  Heninger, Ahmad Moghimi, and Yuval Yarom.
\newblock {Cachequote: Efficiently recovering long-term secrets of SGX EPID via
  cache attacks}.
\newblock In {\em CHES}, 2018.

\bibitem{Diao2016}
Wenrui Diao, Xiangyu Liu, Zhou Li, and Kehuan Zhang.
\newblock {No Pardon for the Interruption: New Inference Attacks on Android
  Through Interrupt Timing Analysis}.
\newblock In {\em S\&P}, 2016.

\bibitem{Dixon2017}
Lizzie Dixon.
\newblock {Breaking KASLR with perf}, 2017.
\newblock URL: \url{https://blog.lizzie.io/kaslr-and-perf.html}.

\bibitem{DomasMovfuscator2015}
Christopher Domas.
\newblock {M/o/Vfuscator}, 2015.
\newblock URL: \url{https://github.com/xoreaxeaxeax/movfuscator}.

\bibitem{Doychev2013CacheAudit}
Goran Doychev, Dominik Feld, Boris Kopf, Laurent Mauborgne, and Jan Reineke.
\newblock {CacheAudit: A Tool for the Static Analysis of Cache Side Channels}.
\newblock In {\em USENIX Security Symposium}, 2013.

\bibitem{Electron2022Apps}
{Electron}.
\newblock {Electron Apps}, 2022.
\newblock URL: \url{https://www.electronjs.org/apps}.

\bibitem{Electron2022ElectronInternals}
{Electron JS}.
\newblock {Electron Internals: Building Chromium as a Library}, 2022.
\newblock URL:
  \url{https://www.electronjs.org/blog/electron-internals-building-chromium-as-a-library}.

\bibitem{fu2017lapd}
Yangchun Fu, Erick Bauman, Raul Quinonez, and Zhiqiang Lin.
\newblock {SGX-LAPD: Thwarting Controlled Side Channel Attacks via Enclave
  Verifiable Page Faults}.
\newblock In {\em RAID}, 2017.

\bibitem{Garcia2017constant}
Cesar~Pereida Garc{\'\i}a and Billy~Bob Brumley.
\newblock {Constant-Time Callees with Variable-Time Callers}.
\newblock In {\em {USENIX} Security Symposium}, 2017.

\bibitem{Gorman2004}
Mel Gorman.
\newblock {\em {Understanding the Linux Virtual Memory Manager}}.
\newblock Prentice Hall Upper Saddle River, 2004.

\bibitem{Gotzfried2017}
Johannes G{\"o}tzfried, Moritz Eckert, Sebastian Schinzel, and Tilo M{\"u}ller.
\newblock {Cache Attacks on Intel SGX}.
\newblock In {\em {EuroSec}}, 2017.

\bibitem{Gras2017aslr}
Ben Gras, Kaveh Razavi, Erik Bosman, Herbert Bos, and Cristiano Giuffrida.
\newblock {ASLR on the Line: Practical Cache Attacks on the MMU.}
\newblock In {\em {NDSS}}, 2017.

\bibitem{Gruss2015dedup}
Daniel Gruss, David Bidner, and Stefan Mangard.
\newblock {Practical Memory Deduplication Attacks in Sandboxed JavaScript}.
\newblock In {\em {ESORICS}}, 2015.

\bibitem{Gruss2019page}
Daniel Gruss, Erik Kraft, Trishita Tiwari, Michael Schwarz, Ari Trachtenberg,
  Jason Hennessey, Alex Ionescu, and Anders Fogh.
\newblock {Page Cache Attacks}.
\newblock In {\em CCS}, 2019.

\bibitem{Gruss2016Prefetch}
Daniel Gruss, Cl\'{e}mentine Maurice, Anders Fogh, Moritz Lipp, and Stefan
  Mangard.
\newblock {Prefetch Side-Channel Attacks: Bypassing SMAP and Kernel ASLR}.
\newblock In {\em CCS}, 2016.

\bibitem{Gruss2015Template}
Daniel Gruss, Raphael Spreitzer, and Stefan Mangard.
\newblock {Cache Template Attacks: Automating Attacks on Inclusive Last-Level
  Caches}.
\newblock In {\em USENIX Security Symposium}, 2015.

\bibitem{Gullasch2011}
David Gullasch, Endre Bangerter, and Stephan Krenn.
\newblock {Cache Games -- Bringing Access-Based Cache Attacks on AES to
  Practice}.
\newblock In {\em S\&P}, 2011.

\bibitem{Gulmezoglu2016cross}
Berk Gulmezoglu, Mehmet~Sinan Inci, Gorka Irazoqui, Thomas Eisenbarth, and Berk
  Sunar.
\newblock {Cross-VM cache attacks on AES}.
\newblock {\em IEEE Transactions on Multi-Scale Computing Systems},
  2(3):211--222, 2016.

\bibitem{Guelmezoglu2015}
Berk G\"{u}lmezo\u{g}lu, Mehmet~Sinan Inci, Thomas Eisenbarth, and Berk Sunar.
\newblock {A Faster and More Realistic Flush+Reload Attack on AES}.
\newblock In {\em {COSADE}}, 2015.

\bibitem{HaloLinux2022Page}
{halolinux}.
\newblock {Page Cache Readahead}, 2022.
\newblock URL:
  \url{https://www.halolinux.us/kernel-architecture/page-cache-readahead.html}.

\bibitem{Harnik2010}
Danny Harnik, Benny Pinkas, and Alexandra Shulman-Peleg.
\newblock {Side channels in cloud services, the case of deduplication in cloud
  storage}.
\newblock {\em IEEE Security \& Privacy}, (6), 2010.

\bibitem{Hund2013}
Ralf Hund, Carsten Willems, and Thorsten Holz.
\newblock {Practical Timing Side Channel Attacks against Kernel Space ASLR}.
\newblock In {\em {S\&P}}, 2013.

\bibitem{Inci2015}
Mehmet~Sinan Inci, Berk Gulmezoglu, Gorka Irazoqui, Thomas Eisenbarth, and Berk
  Sunar.
\newblock {Seriously, get off my cloud! Cross-VM RSA Key Recovery in a Public
  Cloud}.
\newblock {\em Cryptology ePrint Archive, Report 2015/898}, 2015.

\bibitem{Inci2016}
Mehmet~Sinan Inci, Berk Gulmezoglu, Gorka Irazoqui, Thomas Eisenbarth, and Berk
  Sunar.
\newblock {Cache Attacks Enable Bulk Key Recovery on the Cloud}.
\newblock In {\em CHES}, 2016.

\bibitem{Jiang2005}
Song Jiang, Feng Chen, and Xiaodong Zhang.
\newblock {CLOCK-Pro: An Effective Improvement of the CLOCK Replacement.}
\newblock In {\em {USENIX ATC}}, 2005.

\bibitem{Moser2006Optimizing}
{John Richard Moser}.
\newblock {Optimizing Linker Load Times}, 2006.
\newblock URL: \url{https://lwn.net/Articles/192624/}.

\bibitem{Corbet2019Fixing}
{Jonathan Corbet}.
\newblock {Fixing page-cache side channels, second attempt}, 2019.
\newblock URL: \url{https://lwn.net/Articles/778437/}.

\bibitem{Joye2002Montgomery}
Marc Joye and Sung-Ming Yen.
\newblock {The Montgomery powering ladder}.
\newblock In {\em CHES}, 2002.

\bibitem{Keelveedhi2013dupless}
Sriram Keelveedhi, Mihir Bellare, and Thomas Ristenpart.
\newblock {DupLESS: Server-Aided Encryption for Deduplicated Storage}.
\newblock In {\em USENIX Security Symposium}, 2013.

\bibitem{Kim2021Breaking}
Taehun Kim, Taehyun Kim, and Youngjoo Shin.
\newblock {Breaking KASLR Using Memory Deduplication in Virtualized
  Environments}.
\newblock {\em Electronics}, 10(17), 2021.

\bibitem{Kocher2019}
Paul Kocher, Jann Horn, Anders Fogh, Daniel Genkin, Daniel Gruss, Werner Haas,
  Mike Hamburg, Moritz Lipp, Stefan Mangard, Thomas Prescher, Michael Schwarz,
  and Yuval Yarom.
\newblock {Spectre Attacks: Exploiting Speculative Execution}.
\newblock In {\em S\&P}, 2019.

\bibitem{Kocher1996}
Paul~C. Kocher.
\newblock {Timing Attacks on Implementations of Diffe-Hellman, RSA, DSS, and
  Other Systems}.
\newblock In {\em CRYPTO}, 1996.

\bibitem{Koschel2020}
Jakob Koschel, Cristiano Giuffrida, Herbert Bos, and Kaveh Razavi.
\newblock {TagBleed: Breaking KASLR on the Isolated Kernel Address Space Using
  Tagged TLBs}.
\newblock In {\em {EuroS\&P}}, 2020.

\bibitem{Lawson2009}
Nate Lawson.
\newblock {Side channel attacks on cryptographic software}.
\newblock {\em IEEE Security \& Privacy}, 7(6):65--68, 2009.

\bibitem{Li2020SCNET}
Guanlin Li, Chang Liu, Han Yu, Yanhong Fan, Libang Zhang, Zongyue Wang, and
  Meiqin Wang.
\newblock {SCNet: A Neural Network for Automated Side-Channel Attack}.
\newblock {\em arXiv:2008.00476}, 2020.

\bibitem{Lipp2022amd}
Moritz Lipp, Daniel Gruss, and Michael Schwarz.
\newblock {AMD Prefetch Attacks through Power and Time}.
\newblock In {\em USENIX Security Symposium}, 2022.

\bibitem{Lipp2016}
Moritz {Lipp}, Daniel {Gruss}, Raphael {Spreitzer}, Clémentine Maurice, and
  Stefan {Mangard}.
\newblock {ARMageddon: Cache Attacks on Mobile Devices}.
\newblock In {\em USENIX Security Symposium}, 2016.

\bibitem{Maurice2017Hello}
Clémentine Maurice, Manuel Weber, Michael Schwarz, Lukas Giner, Daniel Gruss,
  Carlo Alberto~Boano, Stefan Mangard, and Kay R{\"o}mer.
\newblock {Hello from the Other Side: SSH over Robust Cache Covert Channels in
  the Cloud}.
\newblock In {\em NDSS}, 2017.

\bibitem{Medwed2008Template}
Marcel Medwed and Elisabeth Oswald.
\newblock {Template attacks on ECDSA}.
\newblock In {\em WISA}. Springer, 2008.

\bibitem{memcached_website}
{Memcached}.
\newblock {memcached - a distributed memory object caching system}, 2020.
\newblock URL: \url{https://memcached.org/}.

\bibitem{Moghimi2017}
Ahmad Moghimi, Gorka Irazoqui, and Thomas Eisenbarth.
\newblock {CacheZoom: How SGX amplifies the power of cache attacks}.
\newblock In {\em {CHES}}, 2017.

\bibitem{InnoDB_PhysicalStructure}
{MySQL}.
\newblock {The Physical Structure of an InnoDB Index}, 2020.
\newblock URL:
  \url{https://dev.mysql.com/doc/refman/8.0/en/innodb-physical-structure.html}.

\bibitem{Naghibijouybari2018}
Hoda Naghibijouybari, Ajaya Neupane, Zhiyun Qian, and Nael Abu-Ghazaleh.
\newblock {Rendered Insecure: GPU Side Channel Attacks are Practical}.
\newblock In {\em {CCS}}, 2018.

\bibitem{nginx}
{nginx}.
\newblock {Advanced Load Balancer, Web Server, \& Reverse Proxy - NGINX}, 2021.
\newblock URL: \url{https://www.nginx.com/}.

\bibitem{nxmnpg2022Manual}
{nxmnpg.lemoda}.
\newblock {Manual Pages - LD.LLD}, 2022.
\newblock URL: \url{https://nxmnpg.lemoda.net/1/ld.lld}.

\bibitem{Oren2015}
Yossef Oren, Vasileios~P Kemerlis, Simha Sethumadhavan, and Angelos~D
  Keromytis.
\newblock {The Spy in the Sandbox: Practical Cache Attacks in JavaScript and
  their Implications}.
\newblock In {\em CCS}, 2015.

\bibitem{Osvik2006}
Dag~Arne Osvik, Adi Shamir, and Eran Tromer.
\newblock {Cache Attacks and Countermeasures: the Case of AES}.
\newblock In {\em CT-RSA}, 2006.

\bibitem{Paccagnella2021lotr}
Riccardo Paccagnella, Licheng Luo, and Christopher~W Fletcher.
\newblock {Lord of the Ring(s): Side Channel Attacks on the {CPU} On-Chip Ring
  Interconnect Are Practical}.
\newblock In {\em USENIX Security Symposium}, 2021.

\bibitem{Page2002}
Dan Page.
\newblock {Theoretical Use of Cache Memory as a Cryptanalytic Side-Channel}.
\newblock {\em Cryptology ePrint Archive, Report 2002/169}, 2002.

\bibitem{Page2006A}
Dan Page.
\newblock {A note on side-channels resulting from dynamic compilation}.
\newblock {\em Cryptology ePrint archive, Report 2006/349}, 2006.

\bibitem{Percival2005}
Colin Percival.
\newblock {Cache Missing for Fun and Profit}.
\newblock In {\em BSDCan}, 2005.

\bibitem{Pereida2016Make}
Cesar Pereida~Garc{\'\i}a, Billy~Bob Brumley, and Yuval Yarom.
\newblock {Make Sure DSA Signing Exponentiations Really Are Constant-Time}.
\newblock In {\em CCS}, 2016.

\bibitem{Rane2015Raccoon}
Ashay Rane, Calvin Lin, and Mohit Tiwari.
\newblock {Raccoon: Closing Digital Side-Channels through Obfuscated
  Execution}.
\newblock In {\em USENIX Security Symposium}, 2015.

\bibitem{Razavi2016}
Kaveh Razavi, Ben Gras, Erik Bosman, Bart Preneel, Cristiano Giuffrida, and
  Herbert Bos.
\newblock {Flip Feng Shui: Hammering a Needle in the Software Stack}.
\newblock In {\em USENIX Security Symposium}, 2016.

\bibitem{Rechberger2004Practical}
Christian Rechberger and Elisabeth Oswald.
\newblock {Practical template attacks}.
\newblock In {\em WISA}, 2004.

\bibitem{Redis2013MemtierBenchmark}
{Redis}.
\newblock {memtier\_benchmark: A High-Throughput Benchmarking Tool for Redis \&
  Memcached}, 2013.
\newblock URL:
  \url{https://redis.com/blog/memtier_benchmark-a-high-throughput-benchmarking-tool-for-redis-memcached}.

\bibitem{Ristenpart2009}
Thomas Ristenpart, Eran Tromer, Hovav Shacham, and Stefan Savage.
\newblock {Hey, You, Get Off of My Cloud: Exploring Information Leakage in
  Third-Party Compute Clouds}.
\newblock In {\em CCS}, 2009.

\bibitem{Roettger2020ridl}
Stephen R{\"o}ttger.
\newblock {Escaping the Chrome Sandbox with RIDL}, 2020.
\newblock URL:
  \url{https://googleprojectzero.blogspot.com/2020/02/escaping-chrome-sandbox-with-ridl.html}.

\bibitem{Ueyama2019lld}
{Rui Ueyama}.
\newblock {lld: A Fast, Simple and Portable Linker}, 2017.
\newblock URL: \url{https://llvm.org/devmtg/2017-10/slides/Ueyama-lld.pdf}.

\bibitem{Russinovich2012}
Mark~E Russinovich, David~A Solomon, and Alex Ionescu.
\newblock {\em {Windows internals}}.
\newblock Pearson Education, 2012.

\bibitem{Saileshwar2021Streamline}
Gururaj Saileshwar, Christopher~W Fletcher, and Moinuddin Qureshi.
\newblock {Streamline: a fast, flushless cache covert-channel attack by
  enabling asynchronous collusion}.
\newblock In {\em ASPLOS}, 2021.

\bibitem{Schwarz2017MGX}
Michael Schwarz, Daniel Gruss, Samuel Weiser, Clémentine Maurice, and Stefan
  Mangard.
\newblock {Malware Guard Extension: Using SGX to Conceal Cache Attacks}.
\newblock In {\em DIMVA}, 2017.

\bibitem{Schwarz2018jstemplate}
Michael Schwarz, Florian Lackner, and Daniel Gruss.
\newblock {JavaScript Template Attacks: Automatically Inferring Host
  Information for Targeted Exploits}.
\newblock In {\em NDSS}, 2019.

\bibitem{SchwarzPteditor}
Michael Schwarz, Moritz Lipp, and Claudio Canella.
\newblock {misc0110/PTEditor: A small library to modify all page-table levels
  of all processes from user space for x86\_64 and ARMv8}, 2018.
\newblock URL: \url{https://github.com/misc0110/PTEditor}.

\bibitem{Schwarz2018KeyDrown}
Michael Schwarz, Moritz Lipp, Daniel Gruss, Samuel Weiser, Clémentine Maurice,
  Raphael Spreitzer, and Stefan Mangard.
\newblock {KeyDrown: Eliminating Software-Based Keystroke Timing Side-Channel
  Attacks}.
\newblock In {\em NDSS}, 2018.

\bibitem{Schwarzl2021Dynamic}
Martin Schwarzl, Pietro Borrello, Andreas Kogler, Kenton Varda, Thomas
  Schuster, Daniel Gruss, and Michael Schwarz.
\newblock Dynamic process isolation.
\newblock {\em arXiv preprint arXiv:2110.04751}, 2021.

\bibitem{Schwarzl2021Specfuscator}
Martin Schwarzl, Claudio Canella, Daniel Gruss, and Michael Schwarz.
\newblock {Specfuscator: Evaluating Branch Removal as a Spectre Mitigation}.
\newblock In {\em {FC}}, 2021.

\bibitem{Schwarzl2022Remote}
Martin Schwarzl, Erik Kraft, Moritz Lipp, and Daniel Gruss.
\newblock {Remote Page Deduplication Attacks}.
\newblock In {\em {NDSS}}, 2022.

\bibitem{Shih2017tsgx}
Ming-Wei Shih, Sangho Lee, Taesoo Kim, and Marcus Peinado.
\newblock {T-SGX: Eradicating controlled-channel attacks against enclave
  programs}.
\newblock In {\em NDSS}, 2017.

\bibitem{Simon2018What}
Laurent Simon, David Chisnall, and Ross Anderson.
\newblock {What you get is what you C: Controlling side effects in mainstream C
  compilers}.
\newblock In {\em EuroS\&P}, 2018.

\bibitem{Song2001}
Dawn~Xiaodong Song, David Wagner, and Xuqing Tian.
\newblock {Timing Analysis of Keystrokes and Timing Attacks on SSH}.
\newblock In {\em USENIX Security Symposium}, 2001.

\bibitem{Statcounter2022Browser}
{statcounter Global Stats}.
\newblock {Browser Market Share Worldwide}, 2022.
\newblock URL: \url{https://gs.statcounter.com/}.

\bibitem{Tsuro2021SpectreJS}
{Stephen R{\"o}ttger and Artur Janc}.
\newblock {A Spectre proof-of-concept for a Spectre-proof web}, 2021.
\newblock URL:
  \url{https://security.googleblog.com/2021/03/a-spectre-proof-of-concept-for-spectre.html}.

\bibitem{Suzaki2011}
Kuniyasu Suzaki, Kengo Iijima, Toshiki Yagi, and Cyrille Artho.
\newblock {Memory Deduplication as a Threat to the Guest OS}.
\newblock In {\em EuroSys}, 2011.

\bibitem{Tsunoo2003}
Yukiyasu Tsunoo, Teruo Saito, and Tomoyasu Suzaki.
\newblock {Cryptanalysis of DES implemented on computers with cache}.
\newblock In {\em {CHES}}, 2003.

\bibitem{Vanbulck2017PTE}
Jo~Van~Bulck, Nico Weichbrodt, R\"udiger Kapitza, Frank Piessens, and Raoul
  Strackx.
\newblock {Telling Your Secrets Without Page Faults: Stealthy Page Table-Based
  Attacks on Enclaved Execution}.
\newblock In {\em USENIX Security Symposium}, 2017.

\bibitem{Van2017Adaptive}
Jeroen Van~Cleemput, Bjorn De~Sutter, and Koen De~Bosschere.
\newblock {Adaptive compiler strategies for mitigating timing side channel
  attacks}.
\newblock {\em TDSC}, 2017.

\bibitem{VanSchaik2018malicious}
Stephan Van~Schaik, Cristiano Giuffrida, Herbert Bos, and Kaveh Razavi.
\newblock {Malicious Management Unit: Why Stopping Cache Attacks in Software is
  Harder Than You Think}.
\newblock In {\em USENIX Security Symposium}, 2018.

\bibitem{VanSchaik2019RIDL}
Stephan van Schaik, Alyssa Milburn, Sebastian {\"o}sterlund, Pietro Frigo,
  Giorgi Maisuradze, Kaveh Razavi, Herbert Bos, and Cristiano Giuffrida.
\newblock {RIDL: Rogue In-flight Data Load}.
\newblock In {\em {S\&P}}, 2019.

\bibitem{Vila2019theory}
Pepe Vila, Boris K{\"o}pf, and Jose Morales.
\newblock {Theory and Practice of Finding Eviction Sets}.
\newblock In {\em S\&P}, 2019.

\bibitem{Intel_DisableHWPrefetcher}
Vish Viswanathan.
\newblock {Disclosure of Hardware Prefetcher Control on Some Intel Processors},
  2014.
\newblock URL:
  \url{https://web.archive.org/web/20160304031330/https://software.intel.com/en-us/articles/disclosure-of-hw-prefetcher-control-on-some-intel-processors}.

\bibitem{Wajahat2019Novel}
Ahsan Wajahat, Azhar Imran, Jahanzaib Latif, Ahsan Nazir, and Anas Bilal.
\newblock {A Novel Approach of Unprivileged Keylogger Detection}.
\newblock In {\em iCoMET}, 2019.

\bibitem{Wang2019Unveiling}
Daimeng Wang, Ajaya Neupane, Zhiyun Qian, Nael Abu-Ghazaleh, Srikanth~V
  Krishnamurthy, Edward~JM Colbert, and Paul Yu.
\newblock {Unveiling your keystrokes: A Cache-based Side-channel Attack on
  Graphics Libraries}.
\newblock In {\em {NDSS}}, 2019.

\bibitem{Wang2017Cached}
Shuai Wang, Pei Wang, Xiao Liu, Danfeng Zhang, and Dinghao Wu.
\newblock $\{$CacheD$\}$: Identifying $\{$Cache-Based$\}$ timing channels in
  production software.
\newblock In {\em USENIX}, 2017.

\bibitem{Github2022ChromeArchive}
{Webnicer Ltd}.
\newblock {chrome-downloads}, 2022.
\newblock URL: \url{https://github.com/webnicer/chrome-downloads/}.

\bibitem{Weiser2018}
Samuel Weiser, Raphael Spreitzer, and Lukas Bodner.
\newblock {Single Trace Attack Against RSA Key Generation in Intel SGX SSL}.
\newblock In {\em {AsiaCCS}}, 2018.

\bibitem{wichelmann2018microwalk}
Jan Wichelmann, Ahmad Moghimi, Thomas Eisenbarth, and Berk Sunar.
\newblock {MicroWalk: A Framework for Finding Side Channels in Binaries}.
\newblock In {\em {ACSAC}}, 2018.

\bibitem{Xiao2013security}
Jidong Xiao, Zhang Xu, Hai Huang, and Haining Wang.
\newblock {Security implications of memory deduplication in a virtualized
  environment}.
\newblock In {\em {International Conference on Dependable Systems and Networks
  (DSN)}}, 2013.

\bibitem{Xu2015controlled}
Yuanzhong Xu, Weidong Cui, and Marcus Peinado.
\newblock {Controlled-Channel Attacks: Deterministic Side Channels for
  Untrusted Operating Systems}.
\newblock In {\em {S\&P}}, 2015.

\bibitem{Xu2011}
Yunjing Xu, Michael Bailey, Farnam Jahanian, Kaustubh Joshi, Matti Hiltunen,
  and Richard Schlichting.
\newblock {An exploration of L2 cache covert channels in virtualized
  environments}.
\newblock In {\em CCSW}, 2011.

\bibitem{Yarom2014Flush}
Yuval Yarom and Katrina Falkner.
\newblock {Flush+Reload: a High Resolution, Low Noise, L3 Cache Side-Channel
  Attack}.
\newblock In {\em USENIX Security Symposium}, 2014.

\bibitem{Yuan2021Automated}
Yuanyuan Yuan, Qi~Pang, and Shuai Wang.
\newblock Automated side channel analysis of media software with manifold
  learning.
\newblock {\em arXiv preprint arXiv:2112.04947}, 2021.

\bibitem{Zhang2009keystroke}
Kehuan Zhang and XiaoFeng Wang.
\newblock {Peeping Tom in the Neighborhood: Keystroke Eavesdropping on
  Multi-User Systems}.
\newblock In {\em {USENIX Security Symposium}}, 2009.

\end{thebibliography}
